\documentclass[
  preprint,
  aps,
  pra,
  amsmath, amssymb,
  superscriptaddress
]{revtex4-2}

\usepackage{graphicx}
\usepackage{newtxtext,newtxmath}
\usepackage{csquotes}
\usepackage{paralist}
\usepackage{xcolor}
\usepackage{etoolbox}
\usepackage{hyperref}

\begin{document}

\title{Acoustic-Driven Surface Cleaning with Millimeter-Sized Bubbles at Translational Resonance}

\author{Yan Jun Lin}
\thanks{These authors contributed equally to this work.}
\affiliation{Department of Biological Sciences, Cornell University, Ithaca, NY 14853, USA}

\author{Zhengyang Liu}
\thanks{These authors contributed equally to this work.}
\affiliation{Department of Biological and Environmental Engineering, Cornell University, Ithaca, NY 14853, USA}

\author{Sunghwan Jung}
\email{sj737@cornell.edu}
\affiliation{Department of Biological and Environmental Engineering, Cornell University, Ithaca, NY 14853, USA}
\date{May 30, 2025}

\begin{abstract}
\noindent Traditional surface cleaning methods often suffer from drawbacks such as chemical harshness, potential for surface damage, and high energy consumption. This study investigates an alternative approach: acoustic-driven surface cleaning using millimeter-sized bubbles excited at low, sub-cavitation frequencies. We identify and characterize a distinct translational resonance of these bubbles, occurring at significantly lower frequencies (e.g., 50 Hz for 1.3 mm diameter bubbles) than the Minnaert resonance for a bubble of the same size. Experiments reveal that at this translational resonance, stationary bubbles exhibit amplified lateral swaying, while bubbles sliding on an inclined surface display pronounced "stop-and-go" dynamics. The theoretical model treats the bubble as a forced, damped harmonic oscillator, where surface tension provides the restoring force and the inertia is dominated by the hydrodynamic added mass of the surrounding fluid. It accurately predicts the observed resonant frequency scaling with bubble size ($\propto R_0^{-3/2}$). Cleaning efficacy, assessed using protein-based artificial soil on glass slides, was improved by approximately 90\% when bubbles were driven at their translational resonant frequency compared to off-resonant frequencies or non-acoustic conditions. These findings demonstrate that leveraging translational resonance enhances bubble-induced shear and agitation, offering an effective and sustainable mechanism for surface cleaning.
\end{abstract}

\maketitle

\section{Introduction}

Traditional methods for cleaning agricultural surfaces often rely on strategies involving harsh chemical agents \cite{gadelha_chemical_2019,mackerer_loperamide_1976}, potentially abrasive processes \cite{martinez-romero_mechanical_2004,li_quantitative_2014}, or energy-intensive techniques \cite{adalja_produce_2018}.  These limitations motivate the exploration of alternative cleaning modalities that promise both high efficacy and preservation of surface integrity. Among emerging approaches, leveraging the dynamics of millimeter-sized bubbles for surface cleaning presents a novel and compelling alternative. Bubble cleaning holds significant promise across various applications, including wastewater treatment \cite{agarwal_principle_2011,qiao_recent_2021}, antifouling \cite{salta_bubbles_2016}, and various methods of surface cleaning \cite{jin_environment-friendly_2022}.

Many methods of surface cleaning utilize ultrasonic frequencies to induce bubble cavitation \cite{howell_ultrasonic_2023,birkin_electrochemical_2015}. Cavitation and shape oscillations at ultrasonic frequencies can enhance cleaning by generating strong shear forces and micro-flows, but these effects can also damage sensitive surfaces or promote microbial growth \cite{yusof_physical_2016,vyas_improved_2020,almalki_ultrasound-assisted_2023,corbett_cleaning_2023}. In cleaning of agricultural surfaces, cavitation can effectively enhance the detachment of bacteria responsible for foodborne illnesses, specifically Listeria monocytogenes and Salmonella Newport \cite{lee_cavitation_2018}. However, cavitation produces intense, erosive forces that can degrade sensitive surfaces \cite{abedini_situ_2023,krella_degradation_2023,ju_experimental_2022}. Ultrasound can also promote microbial growth and biofilm formation under certain conditions \cite{avhad_ultrasound_2015,bochu_influence_2003,sulaiman_ultrasound-assisted_2011,lanchun_research_2003,pitt_ultrasound_2003}. Beyond ultrasonic effects, cavitation introduces localized turbulence and microstreaming, which can increase biofilm formation by enhancing nutrient and oxygen transport to biofilms and thereby promoting bacterial growth \cite{tsagkari_turbulence_2018}. Using millimeter-sized bubbles offers a potential solution, enabling effective cleaning via shear forces while avoiding the drawbacks of cavitation and ultrasound. During both impact and subsequent sliding, millimeter-scale bubbles generate significant shear stress on submerged surfaces. For instance, a bubble with a 550 {\textmu}m radius produces over 300 Pa of shear upon impact (due to rapid changes in thin film curvature) and sustains about 45 Pa during sliding \cite{esmaili_bubble_2019}. These stresses can exceed the critical thresholds for removing bacteria, ranging from 0.03–5 Pa for E. coli B/r and 24–144 Pa for Listeria monocytogenes \cite{owens_inhibition_1987}. Additionally, surface inclination influences the generated shear. Precise experiments tracking the thin liquid film during bubble sliding report 10–50 Pa, with a maximum at 25° inclination for a 0.75 mm bubble \cite{hamidzadeh_thin_2024}. Larger bubbles create higher shear at steeper angles, and an optimal inclination of 22.5° with 0.6 mm bubbles was found to generate enough stress to remove both protein dirt coatings and E. coli \cite{hooshanginejad_effect_2023}. Lastly, it was found that adding a driving frequency at 100 Hz further enhances cleaning, though the mechanism remains unclear \cite{hooshanginejad_cleaning_2022}. 

The application of acoustics to bubble-based cleaning has traditionally emphasized cavitation effects, which is often unsuitable for delicate surfaces due to its potential for material erosion and unintended promotion of microbial growth. In contrast, driving bubbles at lower, sub-cavitation frequencies allows for oscillatory behavior without inducing cavitation or excessive turbulence. Under these conditions, we observe a pronounced enhancement in the translational motion of bubbles at a specific low frequency, an effect strongly indicative of resonance. While the conventional Minnaert resonance, associated with volumetric oscillations, occurs in the kilohertz range for millimeter-scale bubbles, the low-frequency peak in translational response identified here is physically distinct. We refer to this phenomenon as translational resonance. In this study, to overcome the adverse effects of ultrasound and cavitation for surface cleaning, we drove bubbles at sub-cavitation frequencies, much lower than the Minnaert resonance frequency. We identified a sharp translational resonance near 50 Hz for bubbles of $R_0=0.65$ mm, resulting in maximal lateral displacement. This resonance produces pronounced motion both in suspended bubbles and in those interacting with surfaces, revealing a novel regime of oscillatory sliding that can enhance surface cleaning. We demonstrate that the resonance frequency depends on the size of the bubble and explained this dependence by an oscillator model, which aligned with experimental results. By driving the bubbles at frequencies close to the resonance, we achieved enhanced surface cleaning efficacy. The detailed mechanism, involving a stop-and-go sliding motion, was also observed and documented. Taken together, these findings demonstrate that leveraging translational resonance enhances bubble-induced shear and agitation, offering an effective mechanism for surface cleaning with broad applications.

\section{Methods}

\subsection{Experimental Apparatus}

\begin{figure}
    \centering
    \includegraphics[width=.7\textwidth]{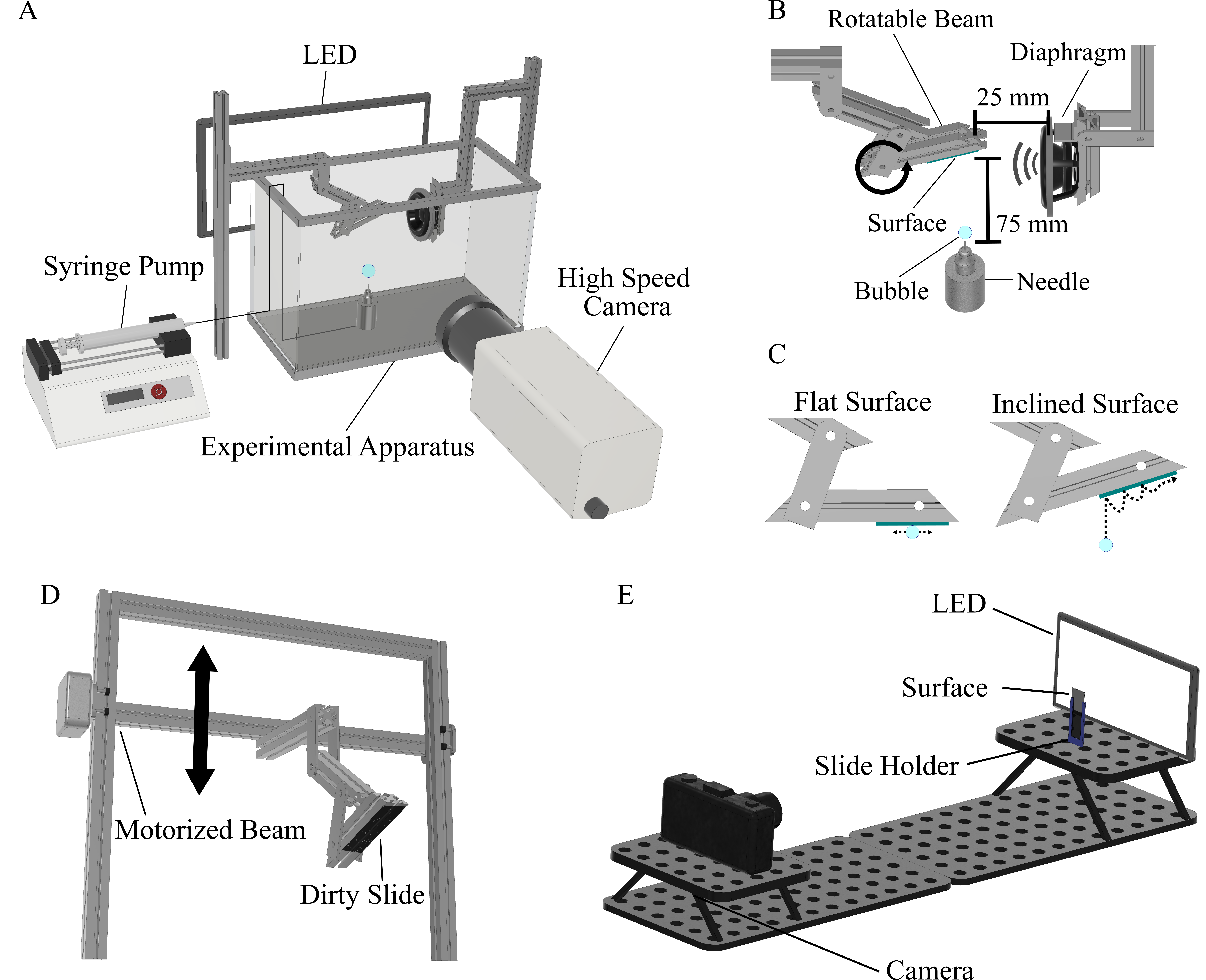} 
    \caption{ (A) Schematic of the apparatus with the essential components. (B) Side view showing the positioning of the surface and speaker relative to the bubble and needle. (C) Illustration of the two surface configurations: flat and inclined. (D) Motorized stage for consistent slide immersion and retraction. (E) Imaging setup for quantifying soil coverage pre- and post-treatment.}
    \label{fig:fig1}
\end{figure}

Experiments were conducted in an open-top glass tank. Millimeter-scale gas bubbles were generated using a 34 gauge needle connected via PVC vinyl tubing to an InfusionONE Single Channel Syringe Pump (New Era Pump Systems Inc.). Two opposing beams were mounted onto an external base plate. One beam supported a mount holding a 75×25 mm glass microscope slide, which served as the test surface (figure~\ref{fig:fig1}A). The opposing beam mounted the acoustic transducer frame, positioned such that the transducer face was approximately 25 mm from the test surface (figure~\ref{fig:fig1}B). The transducer was driven by a Tektronix AFG3101C function generator which provided a sinusoidal driving waveform. 

Videos were captured at 3000 FPS using a Photron FASTCAM NOVA S6 high-speed camera with background illumination from an LED backlight to enhance image contrast. The mean diameter of the bubbles emerging from the 34 gauge needle was 1.3 ± 0.05 mm. This apparatus served as the foundation for multiple experiments, with variations primarily involving the surface inclination (figure~\ref{fig:fig1}C).

\subsection{Dynamics of bubble-mediated cleaning}

The dynamics of low frequency excitation were first investigated using a flat surface configuration (figure~\ref{fig:fig1}C). In this setup, a stationary bubble was suspended beneath a glass slide. This approach minimized confounding external influences, such as variable surface interactions and non-uniform acoustic exposure in sliding bubbles, providing a stable reference compared to more physically complex cases such as bouncing and sliding bubbles. The acoustic field was applied across a frequency range of 5 Hz to 120 Hz (5 Hz intervals) using the standard bubble size (1.3±0.05 mm). Videos of the bubble's response to the acoustic field enabled the assessment of translational oscillation, or swaying. The translational oscillation was characterized by tracking the bubble's centroid position over time. The amplitudes of the bubble’s centroid at various driving frequencies were assessed to identify a peak in response. 

The flat surface configuration was also employed to investigate the dependence of the translational oscillation on bubble size. For these experiments, bubbles of several distinct radii (0.34, 0.46, 0.48, 0.62, 0.67, 0.76, 0.86, 1.33 mm) were generated and suspended beneath the horizontal slide. Each bubble size was tested over a tailored frequency range, with finer increments near experimentally observed peaks in lateral displacement. The input voltage (1 V) was held constant across all experiments. The amplitude of translational oscillation was quantified as a function of driving frequency for each size, characterizing the size-dependent nature of the sub-resonant peak.

To model bubble dynamics in surface cleaning, we employed an inclined surface configuration (figure~\ref{fig:fig1}C). The surface inclination was set at approximately 22°, an angle identified by Hooshanginejad et al. (2023) as yielding maximum shear stress and cleaning effect \cite{hooshanginejad_effect_2023}. In this setup, a bubble detaches from the needle, rises, and bounces upon impact. After the bouncing dampens, the bubble slides along the surface, where the majority of surface cleaning occurs. During experiments, the syringe pump maintained a consistent gas flow rate (5 mL/hr), which allowed each video to isolate a single sliding bubble.  Driving frequencies consistent with previous experiments were applied (5–120 Hz). Five trials per frequency were performed to account for the greater stochasticity arising from the complex bouncing and sliding dynamics on the inclined surface. Centroid tracking provided important trajectory and kinematic data, which gave insight into how acoustic driving can affect surface cleaning. 

\subsection{Acoustic bubble-mediated cleaning of Protein-Based Soil Coating}

A protein-based artificial soil was developed as a model substrate to simulate surface fouling. The artificial soil coating was preferred due to its compositional homogeneity and reproducibility compared to natural soils, which exhibit significant variability in composition, particle size, and adhesion properties.

\begin{table}[h]
    \centering
    \renewcommand{\arraystretch}{1.2} 
    \setlength{\tabcolsep}{10pt} 
    \begin{tabular}{lcccc}
        \hline
        \textbf{Component} & \textbf{Fuller’s Earth} & \textbf{Nigrosin Dye} & \textbf{Water} & \textbf{Milk} \\
        \hline
        \textbf{Amount} & 7.5 g & 0.3 g & 6 mL & 6 mL \\
        \hline
    \end{tabular}
    \caption{ Composition of the artificial dirt mixture used in the study.}
    \label{tab:dirt_composition}
\end{table}

The artificial soil composition is detailed in Table 1 and includes Fuller's Earth (mineral component), nigrosin dye (visual contrast agent), water (solvent), and whole milk (organic component). The selected proportions aim to mimic typical topsoil composition. The solid components were comprised of 7.5 g Fuller's Earth (mineral) and organic matter derived from 6 mL of whole milk. Using a standard solids content of 13\% for whole milk \cite{noauthor_definitions_nodate}, the estimated organic component yields a final mixture containing approximately 90.6\% mineral and 9.4\% organic matter by weight. This composition aligns with the 90:10 mineral to organic matter ratio characteristic of the solid fraction in mineral soils \cite{kalev_chapter_2018}. To ensure coating consistency across samples, component ratios were kept constant by preparing a single large batch prior to spin-coating. 

A total of 30 coated glass slides were prepared (N=5 per condition). The artificial soil mixture was spin-coated onto glass slides at 3000 rpm for 15 seconds. The coated slides were subsequently dried at room temperature for 8 hours, which was empirically optimized to balance adhesion and removal. While insufficient drying led to premature dissolution, excessive drying caused brittleness and fragmentation. The 8 hour period provided adequate adhesion and ensured removal was characteristic of the bubble-mediated surface cleaning rather than submersion. 

Prior to each experiment, the initial state of the soil-coated slide was captured using an imaging setup with LED backlighting in a dark enclosure to generate high contrast between the soil layer and cleared regions (figure~\ref{fig:fig1}E). The experimental apparatus utilized a motorized stage to precisely control the timing and speed of the slide's immersion and retraction (figure~\ref{fig:fig1}D). Cleaning efficacy was evaluated under five experimental conditions, each tested under 5 trials and submerged for 3 minutes. Two conditions involved submersion only: one without acoustic excitation (0 Hz) and one under 50 Hz driving frequency. Bubble-mediated cleaning was performed under a flow rate of 40 mL/hour, tested at 0 Hz, 50 Hz, and 120 Hz. Following the cleaning, the slides were dried for 1 hour to ensure that they were visually free of surface moisture before being imaged using the identical setup for pre-cleaning assessment (figure~\ref{fig:fig1}E).

During processing, the raw image was converted to 8-bit grayscale. A fixed region of interest (ROI) was applied across all images, excluding regions near slide edges to mitigate artifacts from the slide-holder or handling during transfer. Within the defined ROI, Otsu's method was utilized to calculate and apply an optimal threshold, generating a binary mask. Pixels above the threshold were classified as clean, and those below the threshold as soil-covered. This binary classification was preferred over alternative methods that quantified cleaning efficiency based on intensity, since binary classification is better suited to characterize the presence or absence of solid soil particles while minimizing inaccuracies caused by residual dye left on cleaned surfaces, which could confound intensity-based measurements. Cleaning efficacy was calculated as the net increase in clean pixels relative to the initial number of soil-covered pixels within the ROI. All image analysis procedures were implemented using MATLAB R2024a with the Image Processing Toolbox. 

Statistical analysis was conducted using MATLAB R2024a with the Statistics and Machine Learning Toolbox. Bubble cleaning conditions (0 Hz, 50 Hz, 120 Hz), which satisfied homogeneity of variances (Levene’s test, $p > 0.05$), were compared using one-way ANOVA with Tukey-Kramer post hoc testing. Submersion conditions (0 Hz, 50 Hz), also variance-homogeneous, were analyzed separately using Welch's t-test to assess frequency effects under submersion. Bubble and submersion conditions were treated as distinct groups due to their differing physical mechanisms, and thus analyzed independently. 

To evaluate the overall effect of bubble cleaning relative to submersion, data from each method were pooled and compared using Welch’s t-test, which accounts for unequal variances. While ANOVA and related post hoc tests (e.g., Games-Howell) are designed to compare data across multiple groups, their validity depends on the assumption that those groups are statistically comparable, particularly in variance structure. Bubble cleaning and submersion groups differ fundamentally in both mechanism and variance, so comparisons across these groups using ANOVA would be statistically inappropriate. Specifically, bubble cleaning exhibits inherently higher variability due to the diverse trajectories individual bubbles can take due to the complex fluid dynamics that govern their paths. Accordingly, we analyzed each group separately, using Welch’s t-test for method-specific comparisons to ensure statistical validity.

\section{Results}

\subsection{Sub-Resonant Translational Oscillations}

\begin{figure}
    \centering
    \includegraphics[width=.9\textwidth]{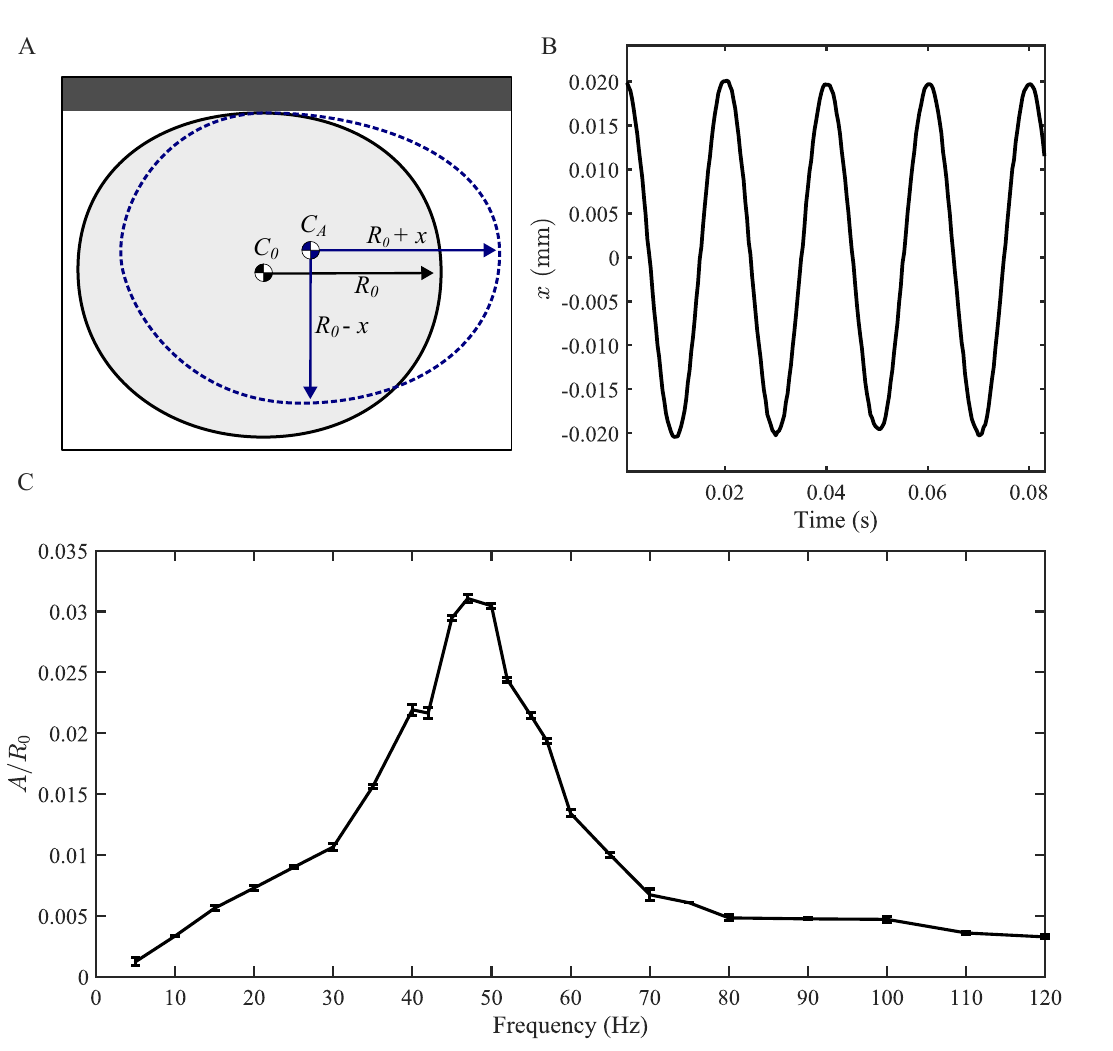} 
    \caption{Translational resonance of an acoustically driven bubble (1.3 mm diameter). (A) Schematic of bubble swaying and deformation under acoustic forcing. A bubble centered at $C_0$ is laterally displaced by an amplitude $x$ to $C_A$, deforming the interface from a sphere to an ellipse. The arrows denote the perturbed semi-axes ($R_0 + x$ and $R_0 - x$). (B) Horizontal displacement of the bubble centroid as a function of time at 50 Hz acoustic excitation, showing periodic swaying in phase with the driving field. (C) Normalized swaying amplitude as a function of driving frequency for a bubble of radius 0.65 mm. A resonance peak is observed at 45 to 50 Hz. Error bars of 1 standard deviation are plotted.    \label{fig:fig2}}
\end{figure}

A bubble with a radius of 0.65 mm was exposed to sinusoidal acoustic excitation spanning 5–120~Hz. In the low-frequency regime, bubbles exhibit a distinct translational mode of oscillation, characterized by lateral displacement of the bubble centroid along the acoustic propagation axis (figure~\ref{fig:fig2}A). The frequency of the bubble’s oscillatory response aligns with the driving frequency at 50 Hz, and the amplitude in displacement characterizes the extent of swaying (figure~\ref{fig:fig2}B). The response amplitudes for this bubble size are most pronounced at 45–50 Hz, indicating resonance at the sharp response peak at this frequency range (figure~\ref{fig:fig2}C).

The flat surface configuration is effective as a preliminary scan for resonance, since there is little variation between trials as shown by the error bars in figure 2C. This eliminates the need for multiple trials per frequency in flat surface experiments, as single trials are sufficient.

\subsection{Theoretical Framework for Translational Resonance in Bubble-Fluid Systems}

\begin{figure}
    \centering
    \includegraphics[width=.9\textwidth]{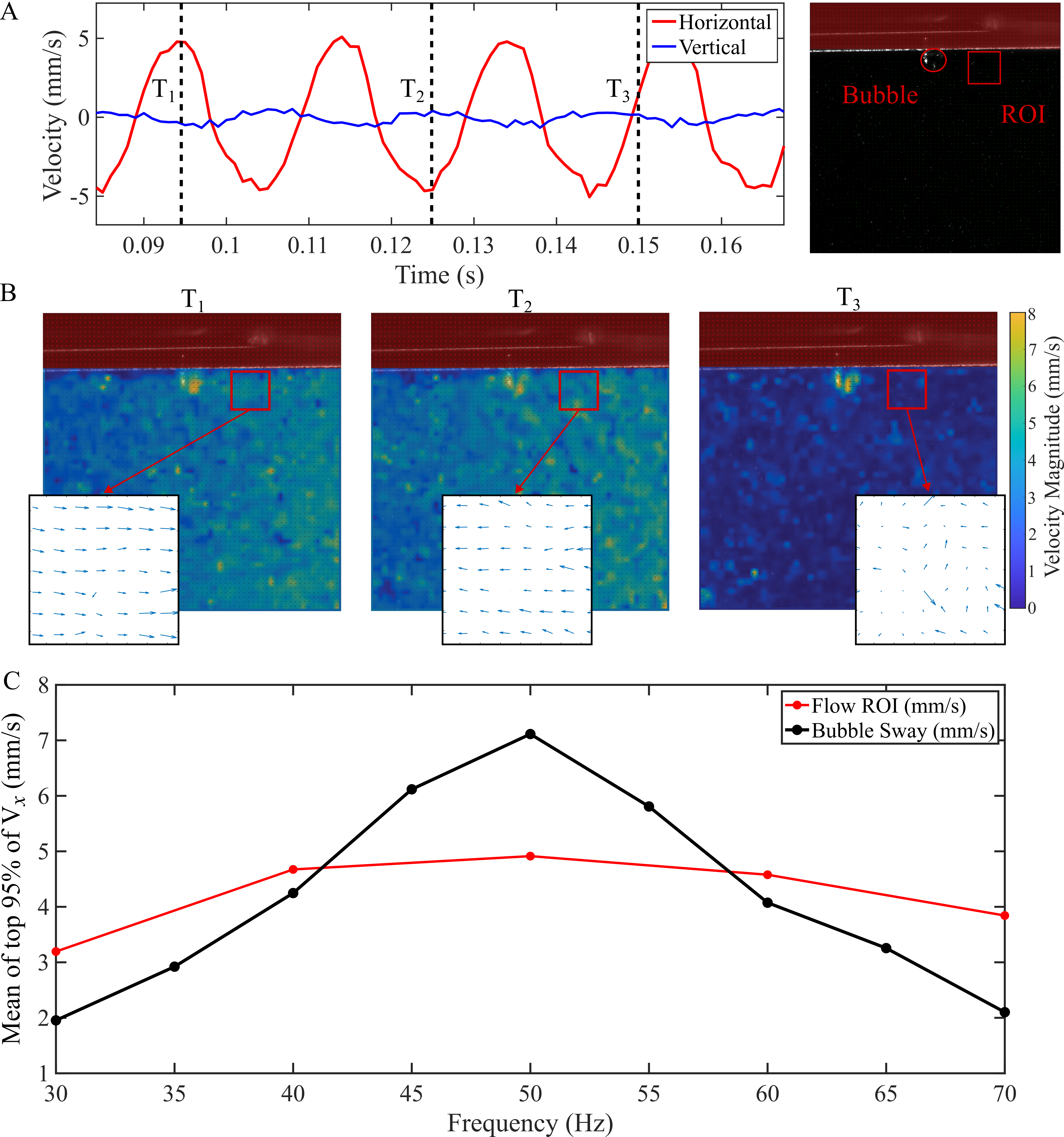} 
    \caption{(A) Time-trace of horizontal and vertical velocities within a fixed region of interest (ROI) between the bubble and the source of the driving frequency. (B) PIV colormap of velocity magnitude at $T_{1}$, $T_{2}$, and $T_{3}$—$T_{1}$ corresponds to maximum pull (rightward), $T_{2}$ to maximum push (leftward), and $T_{3}$ to near‐zero flow. Zoomed‐in vector maps illustrate local flow direction. (C) Mean 95\%-percentile horizontal fluid velocity in the ROI and mean 95\%-percentile bubble swaying velocity versus driving frequency, showing translational amplification surpassing the fluid velocity near 50 Hz.\label{fig:fig3}}
\end{figure}

To elucidate the origin of this resonance behavior, we first characterized the local fluid motion generated by the transducer. As shown in figure~\ref{fig:fig3}A, planar particle image velocimetry (PIV) was used to obtain time-resolved velocity vector maps in the region between the suspended bubble and the transducer. The surface was masked, and a fixed region of interest (ROI) surrounding the bubble was selected for analysis. Within the ROI, the mean horizontal and vertical velocity components were calculated at each time point. Figure~\ref{fig:fig3}B illustrates the bidirectional push–pull effect produced by the diaphragm transducer, indicating the presence of oscillatory fluid movement. The time evolution of the mean horizontal velocities demonstrates a sinusoidal pattern that matches the driving frequency at 50 Hz. 

To evaluate fluid velocity across driving frequencies, we measured the horizontal velocity distribution within the ROI at each frequency. The 95th percentile of x-velocity served as a robust metric for peak fluid velocity (figure~\ref{fig:fig3}C). We utilized the same metric for the swaying bubble to compare the oscillatory motion of the bubble with its surrounding fluid. For the bubble, the same experimental data from the $R_0 = 0.65$ mm bubble from figure~\ref{fig:fig2} was used. 

 While local fluid velocity remained relatively constant across frequencies, the bubble’s swaying velocity showed a sharp resonance peak near 50 Hz, substantially exceeding the local fluid motion at this frequency. Outside the resonance band (~45–55 Hz), bubble velocity was substantially attenuated and dipped below the fluid velocity. This behavior indicates that the bubble is less responsive to the imposed oscillatory flow outside a frequency range of strong amplification, confirming the presence of a translational resonance. By standardizing the driving voltage and measuring both local fluid velocity and bubble centroid velocity, we demonstrate that the observed amplification in bubble swaying arises from an intrinsic resonance phenomenon, not from variations in acoustic input or tracking of the surrounding fluid motion. These findings establish a mechanistic basis for theoretical modeling.  As detailed below, the frequency-dependent amplification is well-described by a forced, damped oscillator framework incorporating added mass, unsteady drag, and surface tension, with the measured fluid velocity serving as the driving term for resonance prediction.

The frequency-dependent amplification of bubble swaying under acoustic excitation is characteristic of a resonant system. We model the bubble as a forced, damped harmonic oscillator. The centroid displacement $x(t)$ obeys:
\begin{equation}
m_{\mathrm{eff}} \frac{d^2x}{dt^2} + b \frac{dx}{dt} + k_{\mathrm{eff}} x = F_{\mathrm{ext}}(t)
\label{eq:eom_concise}
\end{equation}
where $m_{\mathrm{eff}}$ is the effective inertial mass, $b$ is the viscous damping coefficient, $k_{\mathrm{eff}}$ is the effective spring constant from surface tension, and $F_{\mathrm{ext}}(t)$ is the oscillatory acoustic driving force.

For small lateral displacements, the restoring force from surface tension ($\sigma$) arises from changes in the bubble's curvature as it deforms. This can be modeled by considering the bubble deforming slightly from a sphere of initial radius $R_0$ into an ellipsoid, with a small deformation amplitude $X$.This approach is analogous to models developed for deformed bouncing drops, such as by Okumura et al.~\cite{Okumura2003}.

The deformation leads to a difference in curvature ($\Delta\kappa$) between the principal axes of the now ellipsoidal bubble. For small displacements $X$ relative to $R_0$, this curvature difference scales as $\Delta\kappa \sim X/R_0^2$. According to the Young-Laplace equation, this curvature difference results in a pressure difference across the bubble interface of $\Delta p \sim \sigma \Delta\kappa$, which in turn leads to $\Delta p \sim \sigma X/R_0^2$ (figure~\ref{fig:fig2}A). This pressure difference, acting over an effective surface area (which can be considered to scale with $R_0^2$), generates a restoring force $F_s$. Our analysis indicates that this restoring force is approximately proportional to the displacement $X$. Specifically, for small deformations, $F_s \simeq \sigma X$. 

Thus, the effective spring constant $k_{\mathrm{eff}} = F_s/X$ is given by:
\begin{equation}
k_{\mathrm{eff}} \simeq \sigma
\label{eq:keff_concise_revised} 
\end{equation}

The negligible mass of the internal gas means $m_{\mathrm{eff}}$ is dominated by the hydrodynamic added mass of the surrounding liquid accelerated with the bubble. For a sphere of radius $R_0$ in an ideal fluid of density $\rho_f$, this added mass is $m_{\mathrm{eff}} = C_A \rho_f (\frac{4}{3}\pi R_0^3)$, where the added mass coefficient is chosen to be $C_A=1/2$ for a spherical object \cite{lamb_hydrodynamics_1945, jung2021swimming}. Thus, 
\begin{equation}
m_{\mathrm{eff}} = \frac{2\pi}{3} \rho_f R_0^3 \,.
\label{eq:meff_concise}
\end{equation}

The undamped natural frequency is then
\begin{equation}
f_\mathrm{peak}^{(\mathrm{theo})} = \frac{1}{2\pi} \sqrt{\frac{\sigma}{\frac{2\pi}{3} \rho_f R_0^3}} \approx \frac{1}{2\pi} \sqrt{\frac{3 \sigma}{2\pi \rho_f R_0^3}} \,.
\label{eq:f0_concise}
\end{equation}
This is the same as the inverse of the capillary time scale. 
For $R_0 = 0.65 \text{ mm}$, $\sigma = 0.072 \text{ N/m}$, and $\rho_f = 1000 \text{ kg/m}^3$, this gives $f_\mathrm{peak}^{(\mathrm{exp})} \approx 56 \text{ Hz}$, closely matching the observed resonance near $50 \text{ Hz}$.

This model predicts a characteristic size dependence for the natural frequency, $f_\mathrm{peak} \propto R_0^{-3/2}$, indicating that larger bubbles should resonate at lower translational frequencies. The translational resonance is thus principally set by the interplay of surface tension and hydrodynamic inertia, and is weakly damped by viscous effects under typical conditions. We now present experimental results to test this theoretical framework, particularly the predicted scaling with bubble size.

\begin{figure} 
    \centering
    \includegraphics[width=.9\textwidth]{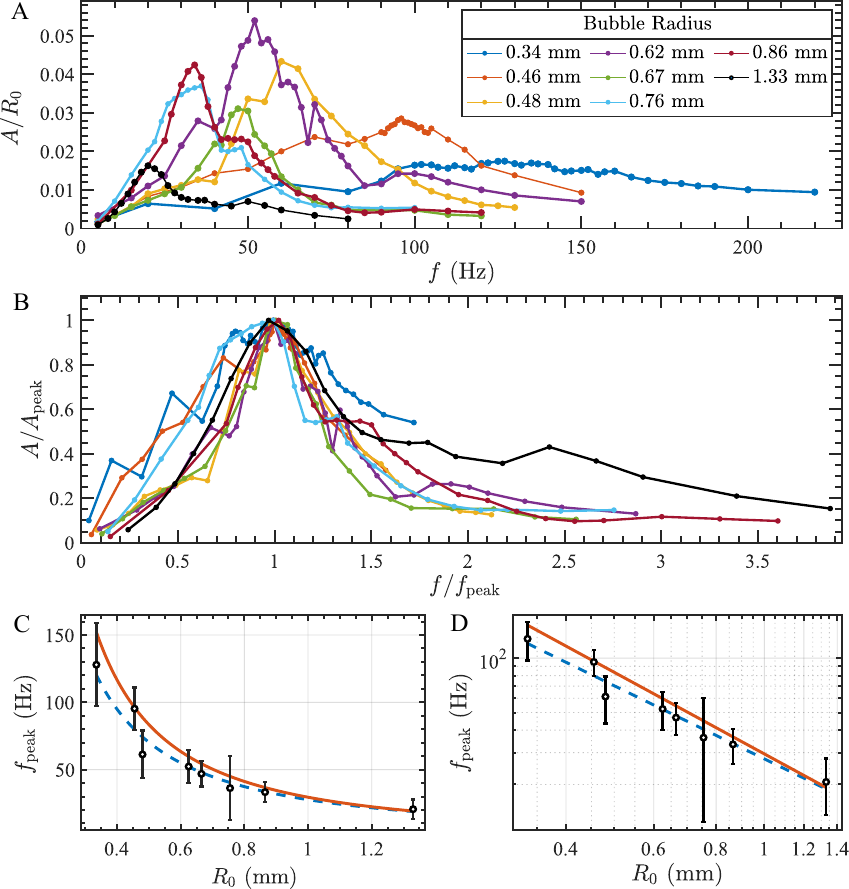} 
    \caption{
    Translational resonance scaling in acoustically driven bubbles.
    (A) Frequency response of normalized oscillation amplitude ($A/R_0$) for bubbles of varying radii.
    (B) Collapsed resonance: normalized amplitude ($A/A_\mathrm{peak}$) versus normalized frequency ($f/f_\mathrm{peak}^{(\mathrm{exp})}$), showing peak alignment across bubble sizes.
    (C) Peak resonant frequency ($f_\mathrm{peak}$) as a function of bubble radius ($R_0$), linear axes. Black: experimental peaks with full width at half maximum (FWHM) error bars. Red: theoretical prediction; blue: experimental power-law fit.
    (D) Log-log scaling of $f_\mathrm{peak}$ vs. $R_0$.  \label{fig:fig4}}
\end{figure}

To experimentally validate the theoretical model, particularly the predicted size dependence of the translational resonant frequency, we characterized the acoustic response of bubbles with radii ranging from approximately $0.33 \text{ mm}$ to $1.33 \text{ mm}$ (figure~\ref{fig:fig4}). The frequency response curves for different bubble sizes, as shown in figure~\ref{fig:fig4}A where normalized sway amplitude ($A/R_0$) is plotted against driving frequency, clearly demonstrate that the peak resonant frequency ($f_\mathrm{peak}^{(\mathrm{exp})}$) systematically decreases as the bubble diameter increases. This qualitative observation aligns with the $f_\mathrm{peak}^{(\mathrm{theo})}$ relationship predicted by Eq. \eqref{eq:f0_concise}.

A quantitative analysis of this size dependence is presented in figure~\ref{fig:fig4}C (linear axes) and (log-log axes), where the experimentally determined $f_\mathrm{peak}^{(\mathrm{exp})}$ values are plotted against bubble radius. The inverse relationship is evident. The log-log plot allows for a direct assessment of the power-law scaling, $f_\mathrm{peak} \propto R_0^{k}$. A linear fit to our experimental data on these logarithmic axes yields a scaling exponent of $k = -1.34$ (figure~\ref{fig:fig4}C).

This experimentally determined exponent of $k = -1.34$ is in close agreement with the theoretical prediction of $k = -1.5$ from Eq. \eqref{eq:f0_concise}. The proximity of the experimental scaling to the theoretical $R^{-3/2}$ dependence provides strong support for the proposed model, affirming that the translational resonance is indeed governed by the balance between surface tension as the restoring force and hydrodynamic inertia (dominated by added mass) as the effective mass. The slight deviation of the experimental exponent from the ideal $-1.5$ may be attributed to idealizations inherent in the model (e.g., the assumptions of the effect spring constant across all sizes and proximity to the surface), experimental uncertainties in determining bubble sizes, or slight non-spherical bubble oscillations. Nevertheless, the observed strong power-law dependence and the alignment of the experimental exponent to the theoretical value robustly validate the core physical mechanisms identified by the theory.

To further validate the observed translational resonance and the underlying harmonic oscillator model, the frequency response curves for different bubble sizes in figure~\ref{fig:fig4}A were replotted in a normalized form. For each bubble size, the oscillation amplitude was normalized by its respective peak amplitude ($A/A_{\mathrm{peak}}$), and the driving frequency was normalized by the experimentally determined peak resonant frequency for that size ($f/f_\mathrm{peak}^{(\mathrm{exp})}$). As shown in figure~\ref{fig:fig4}B, the data from all tested bubble sizes collapse onto a single master curve. This data collapse strongly suggests that the bubble's response is self-similar across the range of sizes investigated when scaled by their characteristic resonant frequencies and corresponding peak amplitudes. It provides compelling evidence for the robustness of the proposed oscillator model and the $R_0^{-3/2}$ scaling of the resonant frequency.

\subsection{Stop-and-Go Sliding and Enhanced Shear from Translational Resonance}

\begin{figure}
    \centering
    \includegraphics[width=.8\textwidth]{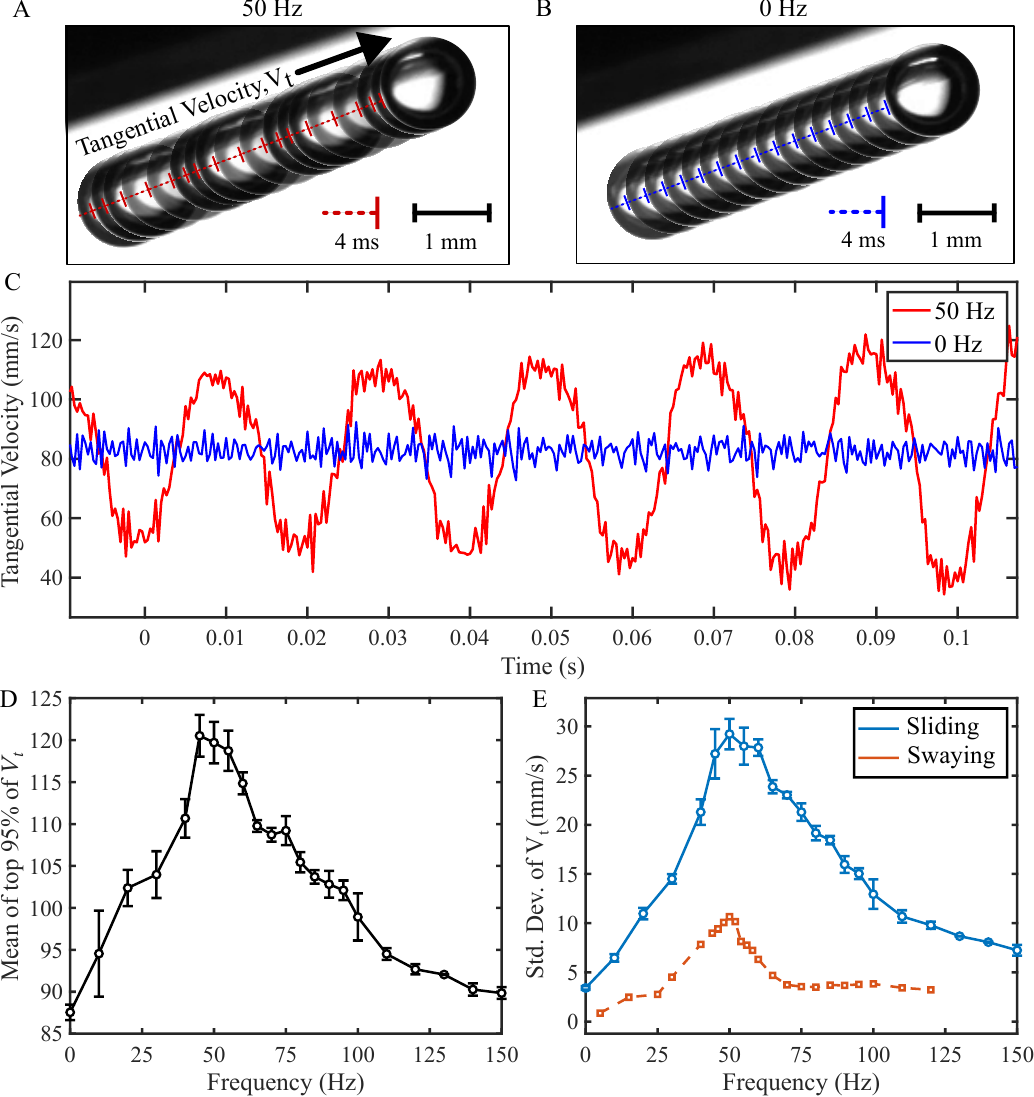} 
    \caption{Translational resonance modulates bubble behavior during inclined sliding. (A) Superimposed time-lapse images of a bubble sliding along 22° inclined glass surface under 50 Hz driving frequency, showing pronounced stop-go motion. (B) Superimposed time-lapse of steady sliding bubble at 0 Hz. (C) Tangential velocity of the sliding bubble as a function of time for both resonant (50 Hz) and static (0 Hz) conditions. (D) 95th percentile of sliding bubble tangential velocity as a function of frequency. (E) Mean SD of sliding velocity and swaying velocity both at 10 volts.\label{fig:fig5}}
\end{figure}

Sliding bubble dynamics under acoustic excitation provide key insight into the mechanism underpinning effective surface cleaning. At 50 Hz, near the resonant frequency identified for stationary swaying for $R_0=0.65$ mm (figure~\ref{fig:fig2}C), bubbles exhibited a pronounced stop-and-go sliding motion (figure~\ref{fig:fig5}A), marked by alternating acceleration and deceleration. In contrast, under static (0 Hz) conditions, bubbles traversed the incline in a steady fashion (figure~\ref{fig:fig5}B).

The velocity profiles reveal the distinctiveness of these regimes. At 50 Hz, the bubble’s tangential velocity oscillates sinusoidally around a mean that closely matches the steady sliding velocity observed at 0 Hz. However, the resonant case is distinguished by large periodic fluctuations; the instantaneous velocity briefly exceeds the steady sliding velocity during acceleration and falls below it during deceleration (figure~\ref{fig:fig5}C). These fluctuations enable the bubble to access transient shear regimes for contaminant removal that are otherwise inaccessible under the steady non-acoustic state. As shown in figure~\ref{fig:fig5}D, the greatest transient velocity occurs between 45 and 55 Hz, which agrees with the resonant frequency established previously (figure~\ref{fig:fig2}C). 

The top 95th percentile velocity quantifies the upper range of velocities achieved by the sliding bubble, providing a robust measure of peak dynamic behavior while minimizing sensitivity to  outliers. This quantifies the high transient velocity reached by the bubble at different driving frequencies, which is essential to cleaning enhancement. In contrast, the standard deviation is more suited for comparing the stationary and sliding bubbles, as it reflects the extent of fluctuating motion rather than absolute speed. The mean standard deviation of velocity of the sliding bubble in comparison to the swaying bubble confirms that the same translational resonance is behind the peak responses in both the sliding and swaying cases (figure~\ref{fig:fig5}E). To enable a direct comparison, both sliding and swaying bubble measurements were performed at the same driving voltage of 10 V. While the swaying bubble exhibited lower translational motion due to their anchored configuration, the peak in their response coincided with that of the sliding bubbles. This congruence underscores a shared underlying inertial–surface tension resonance mechanism in both cases.

\subsection{Surface Cleaning of Artificial Soil-coated Slides}

To evaluate the cleaning efficacy of acoustically driven bubbles, we compared removal outcomes under static (0 Hz), resonant (50 Hz), and off-resonant (120 Hz) excitation, as well as passive submersion controls. Quantitative image analysis (figure~\ref{fig:fig6}A) showed that bubble-mediated cleaning removed 19.3\% of the coated surface at 0 Hz, 33.9\% at 50 Hz, and 17.7\% at 120 Hz, whereas submersion alone resulted in minimal removal: 1.56\% at 0 Hz and 1.42\% at 50 Hz. Visual inspection (figure~\ref{fig:fig6}B) confirmed substantial differences between bubble-treated and submersion-only surfaces. Statistical analysis was conducted separately for bubble cleaning and submersion groups. For bubble cleaning conditions, one-way ANOVA followed by Tukey–Kramer post hoc testing (a = 0.05) identified 50 Hz as yielding significantly greater removal than both 0 Hz and 120 Hz, which were not significantly different from each other. For submersion controls, pairwise t-tests indicated no significant difference between 0 Hz and 50 Hz conditions. An additional pairwise t-test comparing pooled bubble cleaning results to pooled submersion results confirmed statistically distinct groups, as indicated by separate lettering in figure~\ref{fig:fig6}A. Comparison across groups demonstrates clear stratification: all bubble-mediated cleaning conditions yielded significantly greater removal than any submersion condition, and cleaning at 50 Hz was significantly more effective than at 0 Hz or 120 Hz. 

\begin{figure}
    \centering
    \includegraphics[width=.8\textwidth]{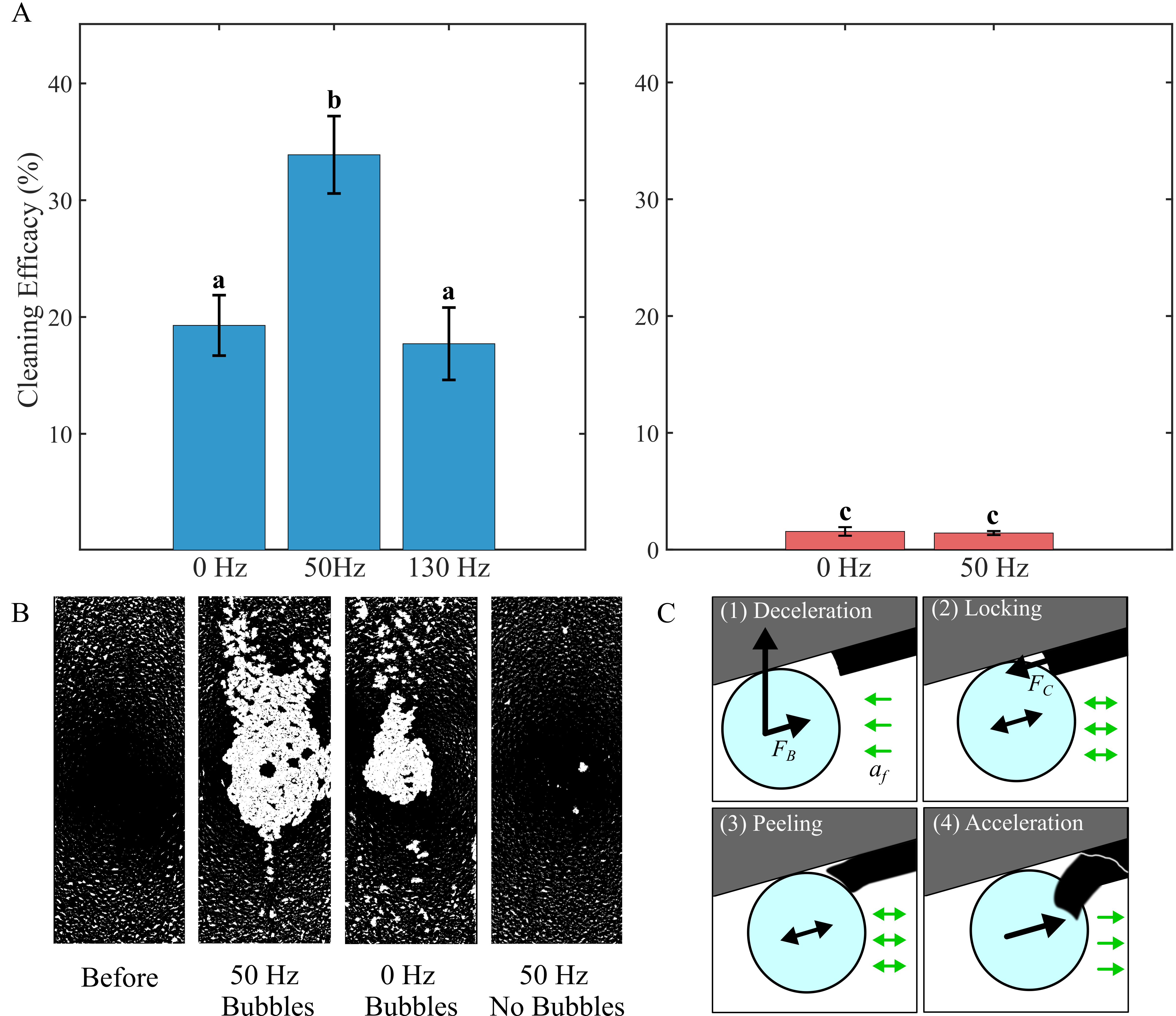} 
    \label{fig:fig6}
\caption{(A) Cleaning efficacy (mean~$\pm$~s.d.) for bubbles interacting with soil-coated slides under static (0~Hz), resonant (50~Hz), and off-resonant (130~Hz) excitation, compared to submersion controls. Statistically distinct groups ($p < 0.05$) are indicated by different lettering. Resonant (50~Hz) bubble cleaning produced the highest removal.
(B) Representative post-cleaning images show substantial removal of the dirt imitation coating with bubble-mediated cleaning at 50~Hz, lower efficacy at 0~Hz, and minimal cleaning by submersion at 50~Hz. 
(C) Schematic of the cleaning mechanism: bubble decelerates, locks with the edge of the surface and peels it away during acceleration with enhanced shear.\label{fig:fig6}}
\end{figure}

These results establish bubble–surface interaction as the dominant mechanism of soil removal, and demonstrate that resonant acoustic excitation near the translational resonant frequency maximizes cleaning efficacy by inducing dynamic sliding and enhanced surface agitation. The nature of this acoustically induced dynamic sliding and its contribution to surface agitation is revealed by a closer examination of the bubble's tangential velocity profile under these resonant conditions. At 50 Hz driving frequency, the bubble's velocity undergoes significant oscillations around a mean value, contrasting with the constant velocity observed during steady sliding at 0 Hz (figure~\ref{fig:fig5}C). This oscillatory kinematic behavior is key to the enhanced cleaning and underpins our proposed mechanism (figure~\ref{fig:fig6}C). 

Initially, oscillatory bubbles work to remove dirt directly from the soiled surface, which opens up cleaned gaps. During the deceleration phase of each oscillation (figure~\ref{fig:fig6}C, panel 1), the bubble's tangential velocity periodically dips well below that of steady sliding. This can facilitate momentary 'locking' against surface contaminants in the soil coating. The oscillations at this lodged position can translate into a localized 'hammering' effect, delivering periodic impulses that mechanically dislodge stubborn contaminants and potentially create clean spots or weakened areas (figure~\ref{fig:fig6}C, panel 2).

The next bubble, or the same bubble in its continued oscillatory path, can then more effectively attack the coating's edge between the cleared area and the remaining dirt. The acceleration phase of the acoustic bubble reaches peak velocities that substantially surpass the average steady sliding speed. This surge generates larger shear stresses precisely at this dirt edge, resulting in a 'peeling' action on the adhered particles (figure~\ref{fig:fig6}C, panels 3 \& 4). The synergy of initial contaminant dislodgement, the creation of attack points at soil edges, intensified shear stresses during acceleration at these edges, and continuous oscillatory mechanical agitation drives the enhanced cleaning efficacy achieved at 50 Hz. An exploration of how such localized mechanical agitation can optimize surface cleaning and broader technological implications are examined further in the Discussion.

\section{Discussion}
Our findings demonstrate that bubble-mediated cleaning efficacy is maximized when bubbles are acoustically driven at their translational resonant frequency, which depends on bubble size. At resonance, bubbles exhibit  amplified motion, leading to significantly enhanced cleaning compared to off-resonant or static conditions. In both the swaying and sliding cases, resonance produces large, phase-synchronized oscillations that maximize local shear and agitation at the surface, leading to highly effective contaminant removal. Off-resonant frequencies result in either steady or weakly modulated motion, both corresponding to diminished cleaning.

The resonance arises because the bubble exhibits an amplified response to localized fluid acceleration produced by the acoustic field. This occurs specifically when the inertia of the bubble and the entrained fluid is balanced by restoring forces from surface tension and contact line pinning. At the translational resonance, even modest acoustic forcing yields large displacements and tangential speeds, transiently increasing shear stress on the surface. The bubble acts as an effective oscillatory “scrubber,” generating bursts of shear that improves cleaning through localized agitation. 

\subsection{Mechanism and Context of Translational Resonance-Driven Surface Cleaning}

Recent computational work by Hooshanginejad et al. (2023) demonstrated that in the absence of acoustic excitation, a bubble sliding along an inclined surface can generate sufficient shear stress at the solid–liquid interface to detach protein aggregates and bacteria, provided the sliding velocity is high enough to surpass established detachment thresholds \cite{hooshanginejad_cleaning_2022}. Their simulations, performed for bubbles of comparable size and under similar geometric conditions to those used here, underscore the feasibility of leveraging bubble-induced surface shear for cleaning applications. Critically, the magnitude of shear stress in their model was strongly dependent on the instantaneous bubble velocity: higher velocities produced greater localized stresses and improved contaminant removal.

Although our study does not directly measure interfacial shear or perform detailed flow simulation due to the complexities of near-field acoustics, our resonance-driven mechanism offers a compelling physical pathway to exceed these detachment thresholds. Driving bubbles at their translational resonance yields transient maxima in bubble velocity and local shear stress that are significantly greater than those achieved by steady, non-resonant sliding (figure~\ref{fig:fig5}D). The numerical modeling demonstrated that the maximum shear stress generated by a sliding bubble (0.6 mm radius) of comparable size to that of the bubble we used in the surface cleaning experiment can readily surpass reported detachment thresholds for a range of microorganisms—including E. coli, Listeria, and Bacillus spores at ~22° inclined surfaces \cite{hooshanginejad_cleaning_2022}. The higher transient bubble velocities in our experimental results may generate shear stresses that are likely to exceed the removal thresholds for even more resilient targets, such as biofilms or stubborn particulate aggregates. This highlights the significant potential of acoustic bubble cleaning driven by translational resonance, particularly as fluctuating, high-magnitude shear has been previously demonstrated to be more effective in surface cleaning than steady forces. Future studies should directly quantify these shear stresses under translational resonant conditions to rigorously evaluate the microbiological cleaning potential of this approach.

At the translational resonant frequency, the bubble’s oscillatory shear is likely to be especially effective at disrupting particle and biofilm adhesion, as multiple studies have shown that unsteady or pulsating flows enhance removal more efficiently than steady shear. For example, Blel et al. (2009) demonstrated that pulsating turbulent flows removed bacterial spores from surfaces more effectively than higher-speed steady flows, attributing the improvement to transient spikes in wall shear stress. Similarly, Ziskind et al. (2000) established that fluctuating shear can weaken adhesive bonds through cyclic fatigue and impulse-driven detachment, mechanisms that steady flows cannot exploit \cite{blel_application_2009,ziskind_particle_2000}. These findings support the hypothesis that the fluctuating shear generated by a resonantly oscillating bubble can surpass adhesion thresholds via repeated high-force bursts rather than relying on time-averaged stress alone. On thinner layers of contaminant, our proposed mechanism for contaminant locking (figure~\ref{fig:fig6}C) likely occurs to a lesser extent, but both the higher transient velocity and the oscillatory nature of the bubble will nevertheless enhance cleaning. 

\subsection{Oscillator Model for Size-Dependent Translational Resonance}

The emergence of translational resonance in our system is governed by a distinct physical mechanism: the bubble behaves as a forced, damped oscillator. In this translational mode, inertial coupling to the surrounding fluid is dominated by the hydrodynamic "added mass," which scales with bubble volume ($m_{\mathrm{eff}} \propto \rho_f R^3$). This inertia is counterbalanced by a restoring force primarily derived from surface tension. The effective spring constant ($k_{\mathrm{eff}}$) can scale as $k_{\mathrm{eff}} \propto \sigma D$ under certain deformation regimes, or more simply as $k_{\mathrm{eff}} \propto \sigma$ if the relevant length scale does not vary strongly with $R_0$. The balance between these terms, with $f_0 \propto \sqrt{k_{\mathrm{eff}}/m_{\mathrm{eff}}}$, leads to amplified lateral displacements when the bubble is driven near its natural frequency.

Our model adopts the simpler $k_{\mathrm{eff}} \sim \sigma$ assumption for tractability and predicts a resonance scaling of $f_0 \propto R_0^{-3/2}$, as given by $f_0 \approx \frac{1}{2\pi} \sqrt{\frac{\sigma}{\frac{2\pi}{3} \rho_f R_0^3}}$. This approach necessarily introduces certain idealizations: for example, treating the bubble as a sphere with a constant added mass coefficient ($C_A=1/2$) and a linear restoring force. In reality, surface proximity can increase both the added mass and the viscous damping, bubble deformation during translation may alter the spring constant, and dynamic contact line pinning introduces nonlinearity to the restoring force. Nevertheless, these simplifications do not obscure the central result: our measured scaling, $f_{\mathrm{peak}} \propto R^{-1.34}$, is in strong agreement with the model prediction, indicating that the dominant physics are well captured.

A key parallel to Minnaert resonance is the size dependence of the resonant frequency. While Minnaert resonance describes radial (volumetric) oscillations that scale as $f_M \propto R^{-1}$, set by the interplay of gas compressibility and fluid inertia \cite{minnaert_musical_1933}, our translational mode follows a steeper $f_0 \propto R^{-3/2}$ scaling governed by surface tension and added mass during lateral motion. Despite operating in different modes of oscillation and being driven by different restoring forces, both resonances demonstrate how the interplay between inertia and restoring forces, such as compressibility for Minnaert resonance or surface tension for the translational resonance, imposes strict and predictable scaling laws on the resonant frequencies of bubble-fluid systems.

The close match between theory and experiment validates the oscillator model and its key assumptions (added mass dominance, surface tension restoring force, and weak damping). The modest deviation from the ideal $-3/2$ exponent likely arises from physical effects not fully incorporated in the simplified model, such as hydrodynamic interactions with the nearby surface, aspherical deformation, and measurement uncertainty. Future theoretical work could address these factors by incorporating more rigorous hydrodynamics, non-linear restoring forces, and direct measurements of damping and contact line mobility. Such refinements would extend the model’s accuracy and range of applicability, providing a more complete framework for predicting and exploiting translational resonance in bubble-mediated processes.

\subsection{Implications and Future Directions}

Translational resonance in acoustically driven bubbles amplifies interfacial shear forces, making surface cleaning more effective. The strong agreement between theory and experiment provides a reliable basis for tuning bubble size and resonance conditions to maximize cleaning performance. This approach is relevant to both biomedical and industrial applications. In medical settings, it may help remove biofilms from devices such as implants and catheters without damaging surfaces. In industry, resonance-driven bubble cleaning could lower chemical and energy use, with potential benefits for food processing, semiconductor cleaning, and filtration. 

Future work should directly quantify shear forces by resolving the thin film and consider how interactions between multiple bubbles and surface features influence cleaning performance. Ultimately, we demonstrate translational resonance can be used to significantly improve bubble-mediated surface cleaning, which holds promising potential over other conventional approaches. We devise a theoretical model that accurately predicts how translational resonance scales with bubble size, and these predictions are well supported by experimental data. These results open the door to new cleaning strategies that are both efficient and adaptable to a range of scientific and industrial needs.

\section*{Ethics} This work did not require ethical approval from a human subject or animal welfare committee. 
\section*{Data accessibility} All data are accessible in https://osf.io/e4aw8/.
\section*{Declaration of AI use.} We have used AI‑assisted technologies in developing Matlab codes to analyze and plot experimental data. 
\section*{Authors’ contributions} Y.L.: 
experiments, formal analysis, funding acquisition, investigation, methodology, supervision, writing—original draft, writing—review and editing;
Z.L.: experiments, formal analysis, theoretical modeling, investigation, methodology, supervision, writing—review and editing;
S.J.: conceptualization, funding acquisition, investigation, methodology, project administration, supervision, writing—review and editing. All authors gave final approval for publication and agreed to be held accountable for the work performed therein.
\section*{Conflict of interest declaration} We declare we have no competing interests. 
\section*{Funding} 
This study is supported by the Einhorn Research Grant and a grant from the Cornell University College of Arts and Sciences and is partially supported by NOAA NA24OARX417C0598-T1-01 and NSF CBET-2401507.
\section*{Acknowledgments}
We thank Justin Chen, Yicong Fu, and Izabella Komperda for their valuable insights and technical guidance throughout the development of this project, as well as the rest of the Jung lab for their support and feedback. 

\bibliographystyle{apsrev4-2}
\bibliography{references}

\begin{thebibliography}{38}%
\makeatletter
\providecommand \@ifxundefined [1]{%
 \@ifx{#1\undefined}
}%
\providecommand \@ifnum [1]{%
 \ifnum #1\expandafter \@firstoftwo
 \else \expandafter \@secondoftwo
 \fi
}%
\providecommand \@ifx [1]{%
 \ifx #1\expandafter \@firstoftwo
 \else \expandafter \@secondoftwo
 \fi
}%
\providecommand \natexlab [1]{#1}%
\providecommand \enquote  [1]{``#1''}%
\providecommand \bibnamefont  [1]{#1}%
\providecommand \bibfnamefont [1]{#1}%
\providecommand \citenamefont [1]{#1}%
\providecommand \href@noop [0]{\@secondoftwo}%
\providecommand \href [0]{\begingroup \@sanitize@url \@href}%
\providecommand \@href[1]{\@@startlink{#1}\@@href}%
\providecommand \@@href[1]{\endgroup#1\@@endlink}%
\providecommand \@sanitize@url [0]{\catcode `\\12\catcode `\$12\catcode `\&12\catcode `\#12\catcode `\^12\catcode `\_12\catcode `\%12\relax}%
\providecommand \@@startlink[1]{}%
\providecommand \@@endlink[0]{}%
\providecommand \url  [0]{\begingroup\@sanitize@url \@url }%
\providecommand \@url [1]{\endgroup\@href {#1}{\urlprefix }}%
\providecommand \urlprefix  [0]{URL }%
\providecommand \Eprint [0]{\href }%
\providecommand \doibase [0]{https://doi.org/}%
\providecommand \selectlanguage [0]{\@gobble}%
\providecommand \bibinfo  [0]{\@secondoftwo}%
\providecommand \bibfield  [0]{\@secondoftwo}%
\providecommand \translation [1]{[#1]}%
\providecommand \BibitemOpen [0]{}%
\providecommand \bibitemStop [0]{}%
\providecommand \bibitemNoStop [0]{.\EOS\space}%
\providecommand \EOS [0]{\spacefactor3000\relax}%
\providecommand \BibitemShut  [1]{\csname bibitem#1\endcsname}%
\let\auto@bib@innerbib\@empty
\bibitem [{\citenamefont {Gadelha}\ \emph {et~al.}(9 17)\citenamefont {Gadelha}, \citenamefont {Allende}, \citenamefont {LpezGlvez}, \citenamefont {Fernndez}, \citenamefont {Gil},\ and\ \citenamefont {Egea}}]{gadelha_chemical_2019}%
  \BibitemOpen
  \bibfield  {author} {\bibinfo {author} {\bibfnamefont {J.}~\bibnamefont {Gadelha}}, \bibinfo {author} {\bibfnamefont {A.}~\bibnamefont {Allende}}, \bibinfo {author} {\bibfnamefont {F.}~\bibnamefont {LpezGlvez}}, \bibinfo {author} {\bibfnamefont {P.}~\bibnamefont {Fernndez}}, \bibinfo {author} {\bibfnamefont {M.}~\bibnamefont {Gil}},\ and\ \bibinfo {author} {\bibfnamefont {J.}~\bibnamefont {Egea}},\ }\href {https://doi.org/10.2903/j.efsa.2019.e170913} {\bibfield  {journal} {\bibinfo  {journal} {{EFSA} Journal}\ }\textbf {\bibinfo {volume} {17}},\ \bibinfo {pages} {e170913} (\bibinfo {year} {2019-09-17})}\BibitemShut {NoStop}%
\bibitem [{\citenamefont {Mackerer}\ \emph {et~al.}(6 10)\citenamefont {Mackerer}, \citenamefont {Clay},\ and\ \citenamefont {Dajani}}]{mackerer_loperamide_1976}%
  \BibitemOpen
  \bibfield  {author} {\bibinfo {author} {\bibfnamefont {C.~R.}\ \bibnamefont {Mackerer}}, \bibinfo {author} {\bibfnamefont {G.~A.}\ \bibnamefont {Clay}},\ and\ \bibinfo {author} {\bibfnamefont {E.~Z.}\ \bibnamefont {Dajani}},\ }\href@noop {} {\bibfield  {journal} {\bibinfo  {journal} {The Journal of Pharmacology and Experimental Therapeutics}\ }\textbf {\bibinfo {volume} {199}},\ \bibinfo {pages} {131} (\bibinfo {year} {1976-10})}\BibitemShut {NoStop}%
\bibitem [{\citenamefont {Martinez-Romero}\ \emph {et~al.}(2004)\citenamefont {Martinez-Romero}, \citenamefont {Serrano}, \citenamefont {Carbonell}, \citenamefont {Castillo}, \citenamefont {Riquelme},\ and\ \citenamefont {Valero}}]{martinez-romero_mechanical_2004}%
  \BibitemOpen
  \bibfield  {author} {\bibinfo {author} {\bibfnamefont {D.}~\bibnamefont {Martinez-Romero}}, \bibinfo {author} {\bibfnamefont {M.}~\bibnamefont {Serrano}}, \bibinfo {author} {\bibfnamefont {A.}~\bibnamefont {Carbonell}}, \bibinfo {author} {\bibfnamefont {S.}~\bibnamefont {Castillo}}, \bibinfo {author} {\bibfnamefont {F.}~\bibnamefont {Riquelme}},\ and\ \bibinfo {author} {\bibfnamefont {D.}~\bibnamefont {Valero}},\ }in\ \href {https://doi.org/10.1007/1-4020-2534-3_8} {\emph {\bibinfo {booktitle} {Production Practices and Quality Assessment of Food Crops: Quality Handling and Evaluation}}},\ \bibinfo {editor} {edited by\ \bibinfo {editor} {\bibfnamefont {R.}~\bibnamefont {Dris}}\ and\ \bibinfo {editor} {\bibfnamefont {S.~M.}\ \bibnamefont {Jain}}}\ (\bibinfo  {publisher} {Springer Netherlands},\ \bibinfo {year} {2004})\ pp.\ \bibinfo {pages} {233--252}\BibitemShut {NoStop}%
\bibitem [{\citenamefont {Li}\ and\ \citenamefont {Thomas}(2 01)}]{li_quantitative_2014}%
  \BibitemOpen
  \bibfield  {author} {\bibinfo {author} {\bibfnamefont {Z.}~\bibnamefont {Li}}\ and\ \bibinfo {author} {\bibfnamefont {C.}~\bibnamefont {Thomas}},\ }\href {https://doi.org/10.1016/j.tifs.2013.12.001} {\bibfield  {journal} {\bibinfo  {journal} {Trends in Food Science \& Technology}\ }\textbf {\bibinfo {volume} {35}},\ \bibinfo {pages} {138} (\bibinfo {year} {2014-02-01})}\BibitemShut {NoStop}%
\bibitem [{\citenamefont {Adalja}\ and\ \citenamefont {Lichtenberg}(1 01)}]{adalja_produce_2018}%
  \BibitemOpen
  \bibfield  {author} {\bibinfo {author} {\bibfnamefont {A.}~\bibnamefont {Adalja}}\ and\ \bibinfo {author} {\bibfnamefont {E.}~\bibnamefont {Lichtenberg}},\ }\href {https://doi.org/10.1016/j.foodpol.2017.10.005} {\bibfield  {journal} {\bibinfo  {journal} {Food Policy}\ }\textbf {\bibinfo {volume} {74}},\ \bibinfo {pages} {23} (\bibinfo {year} {2018-01-01})}\BibitemShut {NoStop}%
\bibitem [{\citenamefont {Agarwal}\ \emph {et~al.}(8 01)\citenamefont {Agarwal}, \citenamefont {Ng},\ and\ \citenamefont {Liu}}]{agarwal_principle_2011}%
  \BibitemOpen
  \bibfield  {author} {\bibinfo {author} {\bibfnamefont {A.}~\bibnamefont {Agarwal}}, \bibinfo {author} {\bibfnamefont {W.~J.}\ \bibnamefont {Ng}},\ and\ \bibinfo {author} {\bibfnamefont {Y.}~\bibnamefont {Liu}},\ }\href {https://doi.org/10.1016/j.chemosphere.2011.05.054} {\bibfield  {journal} {\bibinfo  {journal} {Chemosphere}\ }\textbf {\bibinfo {volume} {84}},\ \bibinfo {pages} {1175} (\bibinfo {year} {2011-08-01})}\BibitemShut {NoStop}%
\bibitem [{\citenamefont {Qiao}\ \emph {et~al.}(3 03)\citenamefont {Qiao}, \citenamefont {Yang}, \citenamefont {Mao}, \citenamefont {Xie}, \citenamefont {Gong}, \citenamefont {Peng}, \citenamefont {Peng}, \citenamefont {Wang}, \citenamefont {Zhang},\ and\ \citenamefont {Zeng}}]{qiao_recent_2021}%
  \BibitemOpen
  \bibfield  {author} {\bibinfo {author} {\bibfnamefont {C.}~\bibnamefont {Qiao}}, \bibinfo {author} {\bibfnamefont {D.}~\bibnamefont {Yang}}, \bibinfo {author} {\bibfnamefont {X.}~\bibnamefont {Mao}}, \bibinfo {author} {\bibfnamefont {L.}~\bibnamefont {Xie}}, \bibinfo {author} {\bibfnamefont {L.}~\bibnamefont {Gong}}, \bibinfo {author} {\bibfnamefont {X.}~\bibnamefont {Peng}}, \bibinfo {author} {\bibfnamefont {Q.}~\bibnamefont {Peng}}, \bibinfo {author} {\bibfnamefont {T.}~\bibnamefont {Wang}}, \bibinfo {author} {\bibfnamefont {H.}~\bibnamefont {Zhang}},\ and\ \bibinfo {author} {\bibfnamefont {H.}~\bibnamefont {Zeng}},\ }\href {https://doi.org/10.1063/5.0040331} {\bibfield  {journal} {\bibinfo  {journal} {Applied Physics Reviews}\ }\textbf {\bibinfo {volume} {8}},\ \bibinfo {pages} {011315} (\bibinfo {year} {2021-03-03})}\BibitemShut {NoStop}%
\bibitem [{\citenamefont {Salta}\ \emph {et~al.}(6 09)\citenamefont {Salta}, \citenamefont {Goodes}, \citenamefont {Maas}, \citenamefont {Dennington}, \citenamefont {Secker},\ and\ \citenamefont {Leighton}}]{salta_bubbles_2016}%
  \BibitemOpen
  \bibfield  {author} {\bibinfo {author} {\bibfnamefont {M.}~\bibnamefont {Salta}}, \bibinfo {author} {\bibfnamefont {L.~R.}\ \bibnamefont {Goodes}}, \bibinfo {author} {\bibfnamefont {B.~J.}\ \bibnamefont {Maas}}, \bibinfo {author} {\bibfnamefont {S.~P.}\ \bibnamefont {Dennington}}, \bibinfo {author} {\bibfnamefont {T.~J.}\ \bibnamefont {Secker}},\ and\ \bibinfo {author} {\bibfnamefont {T.~G.}\ \bibnamefont {Leighton}},\ }\href {https://doi.org/10.1088/2051-672X/4/3/034009} {\bibfield  {journal} {\bibinfo  {journal} {Surface Topography: Metrology and Properties}\ }\textbf {\bibinfo {volume} {4}},\ \bibinfo {pages} {034009} (\bibinfo {year} {2016-09})},\ \bibinfo {note} {publisher: {IOP} Publishing}\BibitemShut {NoStop}%
\bibitem [{\citenamefont {Jin}\ \emph {et~al.}(7 01)\citenamefont {Jin}, \citenamefont {Zhang}, \citenamefont {Cui}, \citenamefont {Sun}, \citenamefont {Gao}, \citenamefont {Pu},\ and\ \citenamefont {Yang}}]{jin_environment-friendly_2022}%
  \BibitemOpen
  \bibfield  {author} {\bibinfo {author} {\bibfnamefont {N.}~\bibnamefont {Jin}}, \bibinfo {author} {\bibfnamefont {F.}~\bibnamefont {Zhang}}, \bibinfo {author} {\bibfnamefont {Y.}~\bibnamefont {Cui}}, \bibinfo {author} {\bibfnamefont {L.}~\bibnamefont {Sun}}, \bibinfo {author} {\bibfnamefont {H.}~\bibnamefont {Gao}}, \bibinfo {author} {\bibfnamefont {Z.}~\bibnamefont {Pu}},\ and\ \bibinfo {author} {\bibfnamefont {W.}~\bibnamefont {Yang}},\ }\href {https://doi.org/10.1016/j.partic.2021.07.008} {\bibfield  {journal} {\bibinfo  {journal} {Particuology}\ }\textbf {\bibinfo {volume} {66}},\ \bibinfo {pages} {1} (\bibinfo {year} {2022-07-01})}\BibitemShut {NoStop}%
\bibitem [{\citenamefont {Howell}\ \emph {et~al.}(3 11)\citenamefont {Howell}, \citenamefont {Ham},\ and\ \citenamefont {Jung}}]{howell_ultrasonic_2023}%
  \BibitemOpen
  \bibfield  {author} {\bibinfo {author} {\bibfnamefont {J.}~\bibnamefont {Howell}}, \bibinfo {author} {\bibfnamefont {E.}~\bibnamefont {Ham}},\ and\ \bibinfo {author} {\bibfnamefont {S.}~\bibnamefont {Jung}},\ }\href {https://doi.org/10.3390/fluids8110291} {\bibfield  {journal} {\bibinfo  {journal} {Fluids}\ }\textbf {\bibinfo {volume} {8}},\ \bibinfo {pages} {291} (\bibinfo {year} {2023-11})},\ \bibinfo {note} {number: 11 Publisher: Multidisciplinary Digital Publishing Institute}\BibitemShut {NoStop}%
\bibitem [{\citenamefont {Birkin}\ \emph {et~al.}(8 12)\citenamefont {Birkin}, \citenamefont {Offin}, \citenamefont {Vian},\ and\ \citenamefont {Leighton}}]{birkin_electrochemical_2015}%
  \BibitemOpen
  \bibfield  {author} {\bibinfo {author} {\bibfnamefont {P.~R.}\ \bibnamefont {Birkin}}, \bibinfo {author} {\bibfnamefont {D.~G.}\ \bibnamefont {Offin}}, \bibinfo {author} {\bibfnamefont {C.~J.~B.}\ \bibnamefont {Vian}},\ and\ \bibinfo {author} {\bibfnamefont {T.~G.}\ \bibnamefont {Leighton}},\ }\href {https://doi.org/10.1039/C5CP02933C} {\bibfield  {journal} {\bibinfo  {journal} {Physical Chemistry Chemical Physics}\ }\textbf {\bibinfo {volume} {17}},\ \bibinfo {pages} {21709} (\bibinfo {year} {2015-08-12})},\ \bibinfo {note} {publisher: The Royal Society of Chemistry}\BibitemShut {NoStop}%
\bibitem [{\citenamefont {Yusof}\ \emph {et~al.}(3 01)\citenamefont {Yusof}, \citenamefont {Babgi}, \citenamefont {Alghamdi}, \citenamefont {Aksu}, \citenamefont {Madhavan},\ and\ \citenamefont {Ashokkumar}}]{yusof_physical_2016}%
  \BibitemOpen
  \bibfield  {author} {\bibinfo {author} {\bibfnamefont {N.~S.~M.}\ \bibnamefont {Yusof}}, \bibinfo {author} {\bibfnamefont {B.}~\bibnamefont {Babgi}}, \bibinfo {author} {\bibfnamefont {Y.}~\bibnamefont {Alghamdi}}, \bibinfo {author} {\bibfnamefont {M.}~\bibnamefont {Aksu}}, \bibinfo {author} {\bibfnamefont {J.}~\bibnamefont {Madhavan}},\ and\ \bibinfo {author} {\bibfnamefont {M.}~\bibnamefont {Ashokkumar}},\ }\href {https://doi.org/10.1016/j.ultsonch.2015.06.013} {\bibfield  {journal} {\bibinfo  {journal} {Ultrasonics Sonochemistry}\ }\textbf {\bibinfo {volume} {29}},\ \bibinfo {pages} {568} (\bibinfo {year} {2016-03-01})}\BibitemShut {NoStop}%
\bibitem [{\citenamefont {Vyas}\ \emph {et~al.}(9 03)\citenamefont {Vyas}, \citenamefont {Wang},\ and\ \citenamefont {Walmsley}}]{vyas_improved_2020}%
  \BibitemOpen
  \bibfield  {author} {\bibinfo {author} {\bibfnamefont {N.}~\bibnamefont {Vyas}}, \bibinfo {author} {\bibfnamefont {Q.}~\bibnamefont {Wang}},\ and\ \bibinfo {author} {\bibfnamefont {A.}~\bibnamefont {Walmsley}},\ }\href {https://doi.org/10.1016/j.ultsonch.2020.105338} {\bibfield  {journal} {\bibinfo  {journal} {Ultrasonics Sonochemistry}\ }\textbf {\bibinfo {volume} {70}},\ \bibinfo {pages} {105338} (\bibinfo {year} {2020-09-03})}\BibitemShut {NoStop}%
\bibitem [{\citenamefont {Almalki}\ and\ \citenamefont {Anand}(3 03)}]{almalki_ultrasound-assisted_2023}%
  \BibitemOpen
  \bibfield  {author} {\bibinfo {author} {\bibfnamefont {T.}~\bibnamefont {Almalki}}\ and\ \bibinfo {author} {\bibfnamefont {S.}~\bibnamefont {Anand}},\ }\href {https://doi.org/10.3390/dairy4010007} {\bibfield  {journal} {\bibinfo  {journal} {Dairy}\ }\textbf {\bibinfo {volume} {4}},\ \bibinfo {pages} {100} (\bibinfo {year} {2023-03})},\ \bibinfo {note} {number: 1 Publisher: Multidisciplinary Digital Publishing Institute}\BibitemShut {NoStop}%
\bibitem [{\citenamefont {Corbett}\ \emph {et~al.}(2 18)\citenamefont {Corbett}, \citenamefont {Wang}, \citenamefont {Smith}, \citenamefont {Liu},\ and\ \citenamefont {Walmsley}}]{corbett_cleaning_2023}%
  \BibitemOpen
  \bibfield  {author} {\bibinfo {author} {\bibfnamefont {C.}~\bibnamefont {Corbett}}, \bibinfo {author} {\bibfnamefont {Q.~.}\ \bibnamefont {Wang}}, \bibinfo {author} {\bibfnamefont {W.}~\bibnamefont {Smith}}, \bibinfo {author} {\bibfnamefont {W.~.}\ \bibnamefont {Liu}},\ and\ \bibinfo {author} {\bibfnamefont {A.~D.}\ \bibnamefont {Walmsley}},\ }\href {https://doi.org/10.1063/5.0173730} {\bibfield  {journal} {\bibinfo  {journal} {Physics of Fluids}\ }\textbf {\bibinfo {volume} {35}},\ \bibinfo {pages} {123335} (\bibinfo {year} {2023-12-18})}\BibitemShut {NoStop}%
\bibitem [{\citenamefont {Lee}\ \emph {et~al.}(9 19)\citenamefont {Lee}, \citenamefont {Eifert}, \citenamefont {Jung},\ and\ \citenamefont {Strawn}}]{lee_cavitation_2018}%
  \BibitemOpen
  \bibfield  {author} {\bibinfo {author} {\bibfnamefont {J.~J.}\ \bibnamefont {Lee}}, \bibinfo {author} {\bibfnamefont {J.~D.}\ \bibnamefont {Eifert}}, \bibinfo {author} {\bibfnamefont {S.}~\bibnamefont {Jung}},\ and\ \bibinfo {author} {\bibfnamefont {L.~K.}\ \bibnamefont {Strawn}},\ }\bibfield  {journal} {\bibinfo  {journal} {Frontiers in Sustainable Food Systems}\ }\textbf {\bibinfo {volume} {2}},\ \href {https://doi.org/10.3389/fsufs.2018.00061} {10.3389/fsufs.2018.00061} (\bibinfo {year} {2018-09-19}),\ \bibinfo {note} {publisher: Frontiers}\BibitemShut {NoStop}%
\bibitem [{\citenamefont {Abedini}\ \emph {et~al.}(1 01)\citenamefont {Abedini}, \citenamefont {Hanke},\ and\ \citenamefont {Reuter}}]{abedini_situ_2023}%
  \BibitemOpen
  \bibfield  {author} {\bibinfo {author} {\bibfnamefont {M.}~\bibnamefont {Abedini}}, \bibinfo {author} {\bibfnamefont {S.}~\bibnamefont {Hanke}},\ and\ \bibinfo {author} {\bibfnamefont {F.}~\bibnamefont {Reuter}},\ }\href {https://doi.org/10.1016/j.ultsonch.2022.106272} {\bibfield  {journal} {\bibinfo  {journal} {Ultrasonics Sonochemistry}\ }\textbf {\bibinfo {volume} {92}},\ \bibinfo {pages} {106272} (\bibinfo {year} {2023-01-01})}\BibitemShut {NoStop}%
\bibitem [{\citenamefont {Krella}(3 02)}]{krella_degradation_2023}%
  \BibitemOpen
  \bibfield  {author} {\bibinfo {author} {\bibfnamefont {A.~K.}\ \bibnamefont {Krella}},\ }\href {https://doi.org/10.3390/ma16052058} {\bibfield  {journal} {\bibinfo  {journal} {Materials}\ }\textbf {\bibinfo {volume} {16}},\ \bibinfo {pages} {2058} (\bibinfo {year} {2023-03-02})}\BibitemShut {NoStop}%
\bibitem [{\citenamefont {Ju}\ and\ \citenamefont {Choi}(2 09)}]{ju_experimental_2022}%
  \BibitemOpen
  \bibfield  {author} {\bibinfo {author} {\bibfnamefont {H.-j.}\ \bibnamefont {Ju}}\ and\ \bibinfo {author} {\bibfnamefont {J.-s.}\ \bibnamefont {Choi}},\ }\href {https://doi.org/10.3390/machines10090793} {\bibfield  {journal} {\bibinfo  {journal} {Machines}\ }\textbf {\bibinfo {volume} {10}},\ \bibinfo {pages} {793} (\bibinfo {year} {2022-09})},\ \bibinfo {note} {number: 9 Publisher: Multidisciplinary Digital Publishing Institute}\BibitemShut {NoStop}%
\bibitem [{\citenamefont {Avhad}\ and\ \citenamefont {Rathod}(1 01)}]{avhad_ultrasound_2015}%
  \BibitemOpen
  \bibfield  {author} {\bibinfo {author} {\bibfnamefont {D.~N.}\ \bibnamefont {Avhad}}\ and\ \bibinfo {author} {\bibfnamefont {V.~K.}\ \bibnamefont {Rathod}},\ }\href {https://doi.org/10.1016/j.ultsonch.2014.04.020} {\bibfield  {journal} {\bibinfo  {journal} {Ultrasonics Sonochemistry}\ }\textbf {\bibinfo {volume} {22}},\ \bibinfo {pages} {257} (\bibinfo {year} {2015-01-01})}\BibitemShut {NoStop}%
\bibitem [{\citenamefont {Bochu}\ \emph {et~al.}(0 01)\citenamefont {Bochu}, \citenamefont {Lanchun}, \citenamefont {Jing}, \citenamefont {Yuanyuan},\ and\ \citenamefont {Yanhong}}]{bochu_influence_2003}%
  \BibitemOpen
  \bibfield  {author} {\bibinfo {author} {\bibfnamefont {W.}~\bibnamefont {Bochu}}, \bibinfo {author} {\bibfnamefont {S.}~\bibnamefont {Lanchun}}, \bibinfo {author} {\bibfnamefont {Z.}~\bibnamefont {Jing}}, \bibinfo {author} {\bibfnamefont {Y.}~\bibnamefont {Yuanyuan}},\ and\ \bibinfo {author} {\bibfnamefont {Y.}~\bibnamefont {Yanhong}},\ }\href {https://doi.org/10.1016/S0927-7765(03)00129-2} {\bibfield  {journal} {\bibinfo  {journal} {Colloids and Surfaces B: Biointerfaces}\ }\textbf {\bibinfo {volume} {32}},\ \bibinfo {pages} {35} (\bibinfo {year} {2003-10-01})}\BibitemShut {NoStop}%
\bibitem [{\citenamefont {Sulaiman}\ \emph {et~al.}(5 15)\citenamefont {Sulaiman}, \citenamefont {Ajit}, \citenamefont {Yunus},\ and\ \citenamefont {Chisti}}]{sulaiman_ultrasound-assisted_2011}%
  \BibitemOpen
  \bibfield  {author} {\bibinfo {author} {\bibfnamefont {A.~Z.}\ \bibnamefont {Sulaiman}}, \bibinfo {author} {\bibfnamefont {A.}~\bibnamefont {Ajit}}, \bibinfo {author} {\bibfnamefont {R.~M.}\ \bibnamefont {Yunus}},\ and\ \bibinfo {author} {\bibfnamefont {Y.}~\bibnamefont {Chisti}},\ }\href {https://doi.org/10.1016/j.bej.2011.01.006} {\bibfield  {journal} {\bibinfo  {journal} {Biochemical Engineering Journal}\ }\textbf {\bibinfo {volume} {54}},\ \bibinfo {pages} {141} (\bibinfo {year} {2011-05-15})}\BibitemShut {NoStop}%
\bibitem [{\citenamefont {Lanchun}\ \emph {et~al.}(7 01)\citenamefont {Lanchun}, \citenamefont {Bochu}, \citenamefont {Zhiming}, \citenamefont {Chuanren}, \citenamefont {Chuanyun},\ and\ \citenamefont {Sakanishi}}]{lanchun_research_2003}%
  \BibitemOpen
  \bibfield  {author} {\bibinfo {author} {\bibfnamefont {S.}~\bibnamefont {Lanchun}}, \bibinfo {author} {\bibfnamefont {W.}~\bibnamefont {Bochu}}, \bibinfo {author} {\bibfnamefont {L.}~\bibnamefont {Zhiming}}, \bibinfo {author} {\bibfnamefont {D.}~\bibnamefont {Chuanren}}, \bibinfo {author} {\bibfnamefont {D.}~\bibnamefont {Chuanyun}},\ and\ \bibinfo {author} {\bibfnamefont {A.}~\bibnamefont {Sakanishi}},\ }\href {https://doi.org/10.1016/S0927-7765(03)00023-7} {\bibfield  {journal} {\bibinfo  {journal} {Colloids and Surfaces B: Biointerfaces}\ }\textbf {\bibinfo {volume} {30}},\ \bibinfo {pages} {43} (\bibinfo {year} {2003-07-01})}\BibitemShut {NoStop}%
\bibitem [{\citenamefont {Pitt}\ and\ \citenamefont {Ross}(2003)}]{pitt_ultrasound_2003}%
  \BibitemOpen
  \bibfield  {author} {\bibinfo {author} {\bibfnamefont {W.~G.}\ \bibnamefont {Pitt}}\ and\ \bibinfo {author} {\bibfnamefont {S.~A.}\ \bibnamefont {Ross}},\ }\href {https://doi.org/10.1021/bp0340685} {\bibfield  {journal} {\bibinfo  {journal} {Biotechnology Progress}\ }\textbf {\bibinfo {volume} {19}},\ \bibinfo {pages} {1038} (\bibinfo {year} {2003})}\BibitemShut {NoStop}%
\bibitem [{\citenamefont {Tsagkari}\ and\ \citenamefont {Sloan}(2018)}]{tsagkari_turbulence_2018}%
  \BibitemOpen
  \bibfield  {author} {\bibinfo {author} {\bibfnamefont {E.}~\bibnamefont {Tsagkari}}\ and\ \bibinfo {author} {\bibfnamefont {W.~T.}\ \bibnamefont {Sloan}},\ }\href {https://doi.org/10.1007/s00449-018-1909-0} {\bibfield  {journal} {\bibinfo  {journal} {Bioprocess and Biosystems Engineering}\ }\textbf {\bibinfo {volume} {41}},\ \bibinfo {pages} {757} (\bibinfo {year} {2018})}\BibitemShut {NoStop}%
\bibitem [{\citenamefont {Esmaili}\ \emph {et~al.}(4 29)\citenamefont {Esmaili}, \citenamefont {Shukla}, \citenamefont {Eifert},\ and\ \citenamefont {Jung}}]{esmaili_bubble_2019}%
  \BibitemOpen
  \bibfield  {author} {\bibinfo {author} {\bibfnamefont {E.}~\bibnamefont {Esmaili}}, \bibinfo {author} {\bibfnamefont {P.}~\bibnamefont {Shukla}}, \bibinfo {author} {\bibfnamefont {J.~D.}\ \bibnamefont {Eifert}},\ and\ \bibinfo {author} {\bibfnamefont {S.}~\bibnamefont {Jung}},\ }\href {https://doi.org/10.1103/PhysRevFluids.4.043603} {\bibfield  {journal} {\bibinfo  {journal} {Physical Review Fluids}\ }\textbf {\bibinfo {volume} {4}},\ \bibinfo {pages} {043603} (\bibinfo {year} {2019-04-29})},\ \bibinfo {note} {publisher: American Physical Society}\BibitemShut {NoStop}%
\bibitem [{\citenamefont {Owens}\ \emph {et~al.}(7 06)\citenamefont {Owens}, \citenamefont {Gingell},\ and\ \citenamefont {Rutter}}]{owens_inhibition_1987}%
  \BibitemOpen
  \bibfield  {author} {\bibinfo {author} {\bibfnamefont {N.~F.}\ \bibnamefont {Owens}}, \bibinfo {author} {\bibfnamefont {D.}~\bibnamefont {Gingell}},\ and\ \bibinfo {author} {\bibfnamefont {P.~R.}\ \bibnamefont {Rutter}},\ }\href {https://doi.org/10.1242/jcs.87.5.667} {\bibfield  {journal} {\bibinfo  {journal} {Journal of Cell Science}\ }\textbf {\bibinfo {volume} {87 ( Pt 5)}},\ \bibinfo {pages} {667} (\bibinfo {year} {1987-06})}\BibitemShut {NoStop}%
\bibitem [{\citenamefont {Hamidzadeh}\ \emph {et~al.}(0 08)\citenamefont {Hamidzadeh}, \citenamefont {Hooshanginejad}, \citenamefont {Huang}, \citenamefont {Phan}, \citenamefont {Jung},\ and\ \citenamefont {Pan}}]{hamidzadeh_thin_2024}%
  \BibitemOpen
  \bibfield  {author} {\bibinfo {author} {\bibfnamefont {F.}~\bibnamefont {Hamidzadeh}}, \bibinfo {author} {\bibfnamefont {A.}~\bibnamefont {Hooshanginejad}}, \bibinfo {author} {\bibfnamefont {K.}~\bibnamefont {Huang}}, \bibinfo {author} {\bibfnamefont {T.~H.}\ \bibnamefont {Phan}}, \bibinfo {author} {\bibfnamefont {S.}~\bibnamefont {Jung}},\ and\ \bibinfo {author} {\bibfnamefont {L.}~\bibnamefont {Pan}},\ }\href {https://doi.org/10.1021/acs.langmuir.4c02901} {\bibfield  {journal} {\bibinfo  {journal} {Langmuir: the {ACS} journal of surfaces and colloids}\ }\textbf {\bibinfo {volume} {40}},\ \bibinfo {pages} {21241} (\bibinfo {year} {2024-10-08})}\BibitemShut {NoStop}%
\bibitem [{\citenamefont {Hooshanginejad}\ \emph {et~al.}(4 28)\citenamefont {Hooshanginejad}, \citenamefont {Sheppard}, \citenamefont {Xu}, \citenamefont {Manyalla}, \citenamefont {Jaicks}, \citenamefont {Esmaili},\ and\ \citenamefont {Jung}}]{hooshanginejad_effect_2023}%
  \BibitemOpen
  \bibfield  {author} {\bibinfo {author} {\bibfnamefont {A.}~\bibnamefont {Hooshanginejad}}, \bibinfo {author} {\bibfnamefont {T.}~\bibnamefont {Sheppard}}, \bibinfo {author} {\bibfnamefont {P.}~\bibnamefont {Xu}}, \bibinfo {author} {\bibfnamefont {J.}~\bibnamefont {Manyalla}}, \bibinfo {author} {\bibfnamefont {J.}~\bibnamefont {Jaicks}}, \bibinfo {author} {\bibfnamefont {E.}~\bibnamefont {Esmaili}},\ and\ \bibinfo {author} {\bibfnamefont {S.}~\bibnamefont {Jung}},\ }\href {https://doi.org/10.1103/PhysRevFluids.8.043602} {\bibfield  {journal} {\bibinfo  {journal} {Physical Review Fluids}\ }\textbf {\bibinfo {volume} {8}},\ \bibinfo {pages} {043602} (\bibinfo {year} {2023-04-28})},\ \bibinfo {note} {publisher: American Physical Society}\BibitemShut {NoStop}%
\bibitem [{\citenamefont {Hooshanginejad}\ \emph {et~al.}(8 03)\citenamefont {Hooshanginejad}, \citenamefont {Sheppard}, \citenamefont {Manyalla}, \citenamefont {Jaicks},\ and\ \citenamefont {Jung}}]{hooshanginejad_cleaning_2022}%
  \BibitemOpen
  \bibfield  {author} {\bibinfo {author} {\bibfnamefont {A.}~\bibnamefont {Hooshanginejad}}, \bibinfo {author} {\bibfnamefont {T.~J.}\ \bibnamefont {Sheppard}}, \bibinfo {author} {\bibfnamefont {J.}~\bibnamefont {Manyalla}}, \bibinfo {author} {\bibfnamefont {J.}~\bibnamefont {Jaicks}},\ and\ \bibinfo {author} {\bibfnamefont {S.}~\bibnamefont {Jung}},\ }in\ \href {https://doi.org/10.1115/FEDSM2022-86897} {\emph {\bibinfo {booktitle} {Volume 2: Multiphase Flow ({MFTC}); Computational Fluid Dynamics ({CFDTC}); Micro and Nano Fluid Dynamics ({MNFDTC})}}}\ (\bibinfo  {publisher} {American Society of Mechanical Engineers},\ \bibinfo {year} {2022-08-03})\ p.\ \bibinfo {pages} {V002T04A007}\BibitemShut {NoStop}%
\bibitem [{noa()}]{noauthor_definitions_nodate}%
  \BibitemOpen
  \href {https://www.idfa.org/definition} {\bibinfo {title} {Definitions}}\BibitemShut {NoStop}%
\bibitem [{\citenamefont {Kalev}\ and\ \citenamefont {Toor}(1 01)}]{kalev_chapter_2018}%
  \BibitemOpen
  \bibfield  {author} {\bibinfo {author} {\bibfnamefont {S.~D.}\ \bibnamefont {Kalev}}\ and\ \bibinfo {author} {\bibfnamefont {G.~S.}\ \bibnamefont {Toor}},\ }in\ \href {https://doi.org/10.1016/B978-0-12-809270-5.00014-5} {\emph {\bibinfo {booktitle} {Green Chemistry}}},\ \bibinfo {editor} {edited by\ \bibinfo {editor} {\bibfnamefont {B.}~\bibnamefont {Trk}}\ and\ \bibinfo {editor} {\bibfnamefont {T.}~\bibnamefont {Dransfield}}}\ (\bibinfo  {publisher} {Elsevier},\ \bibinfo {year} {2018-01-01})\ pp.\ \bibinfo {pages} {339--357}\BibitemShut {NoStop}%
\bibitem [{\citenamefont {Okumura}\ \emph {et~al.}(2003)\citenamefont {Okumura}, \citenamefont {Chevy}, \citenamefont {Richard}, \citenamefont {Qu\'{e}r\'{e}},\ and\ \citenamefont {Clanet}}]{Okumura2003}%
  \BibitemOpen
  \bibfield  {author} {\bibinfo {author} {\bibfnamefont {K.}~\bibnamefont {Okumura}}, \bibinfo {author} {\bibfnamefont {F.}~\bibnamefont {Chevy}}, \bibinfo {author} {\bibfnamefont {D.}~\bibnamefont {Richard}}, \bibinfo {author} {\bibfnamefont {D.}~\bibnamefont {Qu\'{e}r\'{e}}},\ and\ \bibinfo {author} {\bibfnamefont {C.}~\bibnamefont {Clanet}},\ }\href {https://doi.org/10.1209/epl/i2003-00208-y} {\bibfield  {journal} {\bibinfo  {journal} {Europhysics Letters ({EPL})}\ }\textbf {\bibinfo {volume} {62}},\ \bibinfo {pages} {237} (\bibinfo {year} {2003})}\BibitemShut {NoStop}%
\bibitem [{\citenamefont {Lamb}(1945)}]{lamb_hydrodynamics_1945}%
  \BibitemOpen
  \bibfield  {author} {\bibinfo {author} {\bibfnamefont {H.}~\bibnamefont {Lamb}},\ }\href {http://archive.org/details/hydrodynamics00lamb} {\emph {\bibinfo {title} {Hydrodynamics}}}\ (\bibinfo  {publisher} {New York,: Dover publications},\ \bibinfo {year} {1945})\BibitemShut {NoStop}%
\bibitem [{\citenamefont {Jung}(2021)}]{jung2021swimming}%
  \BibitemOpen
  \bibfield  {author} {\bibinfo {author} {\bibfnamefont {S.}~\bibnamefont {Jung}},\ }\href@noop {} {\bibfield  {journal} {\bibinfo  {journal} {Scientific Reports}\ }\textbf {\bibinfo {volume} {11}},\ \bibinfo {pages} {15984} (\bibinfo {year} {2021})}\BibitemShut {NoStop}%
\bibitem [{\citenamefont {Blel}\ \emph {et~al.}(2 01)\citenamefont {Blel}, \citenamefont {Legentilhomme}, \citenamefont {Bnzech}, \citenamefont {Legrand},\ and\ \citenamefont {Le~Gentil-Lelivre}}]{blel_application_2009}%
  \BibitemOpen
  \bibfield  {author} {\bibinfo {author} {\bibfnamefont {W.}~\bibnamefont {Blel}}, \bibinfo {author} {\bibfnamefont {P.}~\bibnamefont {Legentilhomme}}, \bibinfo {author} {\bibfnamefont {T.}~\bibnamefont {Bnzech}}, \bibinfo {author} {\bibfnamefont {J.}~\bibnamefont {Legrand}},\ and\ \bibinfo {author} {\bibfnamefont {C.}~\bibnamefont {Le~Gentil-Lelivre}},\ }\href {https://doi.org/10.1016/j.jfoodeng.2008.07.019} {\bibfield  {journal} {\bibinfo  {journal} {Journal of Food Engineering}\ }\textbf {\bibinfo {volume} {90}},\ \bibinfo {pages} {433} (\bibinfo {year} {2009-02-01})}\BibitemShut {NoStop}%
\bibitem [{\citenamefont {Ziskind}\ \emph {et~al.}(6 01)\citenamefont {Ziskind}, \citenamefont {Fichman},\ and\ \citenamefont {Gutfinger}}]{ziskind_particle_2000}%
  \BibitemOpen
  \bibfield  {author} {\bibinfo {author} {\bibfnamefont {G.}~\bibnamefont {Ziskind}}, \bibinfo {author} {\bibfnamefont {M.}~\bibnamefont {Fichman}},\ and\ \bibinfo {author} {\bibfnamefont {C.}~\bibnamefont {Gutfinger}},\ }\href {https://doi.org/10.1016/S0021-8502(99)00554-6} {\bibfield  {journal} {\bibinfo  {journal} {Journal of Aerosol Science}\ }\textbf {\bibinfo {volume} {31}},\ \bibinfo {pages} {703} (\bibinfo {year} {2000-06-01})}\BibitemShut {NoStop}%
\bibitem [{\citenamefont {Minnaert}(1933)}]{minnaert_musical_1933}%
  \BibitemOpen
  \bibfield  {author} {\bibinfo {author} {\bibfnamefont {M.}~\bibnamefont {Minnaert}},\ }\href {https://doi.org/10.1080/14786443309462277} {\bibfield  {journal} {\bibinfo  {journal} {The London, Edinburgh, and Dublin Philosophical Magazine and Journal of Science}\ }\textbf {\bibinfo {volume} {16}},\ \bibinfo {pages} {235} (\bibinfo {year} {1933})}\BibitemShut {NoStop}%
\end{thebibliography}%

\end{document}